\documentclass{aa}  

\usepackage{graphicx}
\usepackage{txfonts}
\usepackage{epsfig}

\def\araa{ARA\&A}%
\def\apj{ApJ}%
\def\apjl{ApJ}%
\def\apjs{ApJS}

\def\aap{A\&A}%
\def\mnras{MNRAS}%
\def\pasj{PASJ}%
\def\aapr{A\&ARv}
\usepackage{color}
\usepackage{hyperref}

\newcommand{\swj}{SWIFT~J1753.5--0127}

\newcommand{\inte}{\textsl{INTEGRAL}}
\newcommand{\xte}{\textsl{RXTE}}
\newcommand{\swift}{\textsl{Swift}}
\newcommand{\maxi}{\textsl{Maxi}}

\def\ergcms{{\rm erg\,cm^{-2}\,s^{-1}}}
\def\Msun{{\rm M_{\odot}}}

\begin{document}

\title{The origin of seed photons for Comptonization in the black hole binary \swj}
\titlerunning{The origin of seed photons for Comptonization in the black hole binary \swj}



\title{The origin of seed photons for Comptonization in the black hole binary {Swift}~J1753.5--0127}

\author{J.~J.~E. Kajava \inst{1, 2, 3} \and
A. Veledina \inst{4, 5} \and
S. Tsygankov \inst{4, 6} \and
V. Neustroev \inst{3,6}}

\institute{
European Space Astronomy Centre (ESA/ESAC), Science Operations Department, 28691 Villanueva de la Ca\~{n}ada, Madrid, Spain\\
\email{jkajava@sciops.esa.int} \and
Nordic Optical Telescope, Apartado 474, 38700 Santa Cruz de La Palma, Spain \and
Astronomy Research Unit, PO Box 3000,  University of Oulu, 90014  Oulu, Finland \and
Tuorla Observatory, University of Turku, V\"{a}is\"{a}l\"{a}ntie 20,  21500 Piikki\"{o}, Finland \and
Nordita, KTH Royal Institute of Technology and Stockholm University, Roslagstullsbacken 23,  10691 Stockholm, Sweden \and
Finnish Centre for Astronomy with ESO (FINCA), University of Turku, V\"{a}is\"{a}l\"{a}ntie 20,  21500 Piikki\"{o}, Finland
}

\date{Received 10 December 2015 / Accepted 29 March 2016}

\abstract{}
{ 
The black hole binary \swj\ is providing a unique data set to study accretion flows. 
Various investigations of this system and of other black holes have not, however, led to an agreement on the accretion flow geometry or on 
the seed photon source for Comptonization during different stages of X-ray outbursts. We  place constraints on these accretion flow 
properties by studying long-term spectral variations of this source.
}
{ 
We performed phenomenological and self-consistent broad band spectral modeling of \swj\ using quasi-simultaneous archived data 
from \inte/ISGRI, \swift/UVOT/XRT/BAT, \xte/PCA/HEXTE, and \maxi/GSC instruments.
}
{ 
We identify a critical flux limit, $F\!\sim\!1.5 \times 10^{-8}\,\ergcms$, and show that the spectral properties of \swj\ are markedly different above and below this value.
Above the limit, during the outburst peak, the hot medium seems to intercept roughly 50~percent of the disk emission.
Below it, in the outburst tail, the contribution of the disk photons reduces significantly and the entire spectrum from the optical to X-rays can be produced by a synchrotron-self-Compton mechanism.
The long-term variations in the hard X-ray spectra are caused by erratic changes of the electron temperatures in the hot medium. 
Thermal Comptonization models indicate unreasonably low hot medium optical depths during the short incursions into the soft state after 2010, suggesting that non-thermal electrons produce the Comptonized tail in this state.
The soft X-ray excess, likely produced by the accretion disk, shows peculiarly stable temperatures for over an order of magnitude changes in flux.
}
{ 
The long-term spectral trends of \swj\ are likely set by variations of the truncation radius and a formation of a hot, quasi-spherical 
inner flow in the vicinity of the black hole.
In the late outburst stages, at fluxes below the critical limit, the source of seed photons for Comptonization is not the thermal disk, 
but more likely they are produced by non-thermal synchrotron emission within the hot flow near the black hole.
The stability of the soft excess temperature is, however, not consistent with this picture and further investigations are needed to understand its behavior.
}
\keywords{black hole physics -- accretion, accretion disks -- X-rays: binaries}
 

\maketitle

\section{Introduction}

Low-mass X-ray binaries (LMXBs) that consist of a black hole (BH) primary and a faint companion star are ideal laboratories to study accretion flows.
When BH-LMXBs undergo \mbox{X-ray} outbursts the emitted electromagnetic spectrum from radio to \mbox{$\gamma$-rays} is produced by accretion-ejection processes.
Observationally, three main spectral components can be identified: 
(1) a hot, optically thin medium (a corona or a hot inner flow in the vicinity of the BH); 
  (2) a cold, geometrically thin, optically thick accretion disk that reflects some of the hard X-ray emission; and 
    (3)~a~jet producing the radio emission.
There is currently no consensus on the accretion geometry or on the processes that produce the seed photons for Comptonization during X-ray \linebreak{outbursts}.

Black hole binaries exhibit two major spectral states: the hard state and the soft state \citep[see, e.g.,][for review]{Esin97,RM06,DGK07}.
These states can be further divided into intermediate states based on spectral and timing correlations (see, e.g., \citealt{Belloni10}).
The soft-state spectra are characterized by a thermal component peaking at $\sim$1~keV.
This emission is attributed to the optically thick, geometrically thin accretion disk \citep{SS73}. 
The weak high-energy tails detected in this state \citep{McConnell02} are likely produced by Compton up-scattering of the disk photons by mostly non-thermal electrons in an optically thin corona or, alternatively, in some discrete active regions above the disk (see, e.g., \citealt{HMG94}).
Strong reflection features -- in the form of iron line(s) at ${\sim}6.4$~keV and a Compton reflection bump in the hard X-ray band --  
imply a geometry where the cold disk extends very close to the BH (see, e.g., \citealt{ZLS99, GCR99}), probably down to the innermost stable circular orbit (ISCO; see, e.g., \citealt{BPT72}). 

The soft state is typically seen only at relatively high luminosities.
However, as soon as the luminosity drops below some limit, BH-LMXBs make a state transition towards the hard state \citep{DFK10}.
The hard state is characterized by a cut-off power-law X-ray spectrum, peaking at ${\sim}100$~keV, which can be explained by thermal 
Comptonization of low-energy seed \mbox{photons} that are generated in the vicinity of the BH. 
A weaker Compton reflection bump is often seen than in the soft state, indicating that the cold accretion disk recedes from the ISCO \citep{Gilfanov10}.

In the hard state the accretion flow geometry is less certain than in the soft state.
Two competing models are considered possible: (1) a disk corona model, where the cold disk is embedded in a hotter medium \citep{P98}
and may extend down to the ISCO if the corona is outflowing \citep{B99PE}; and (2)~the hot inner flow model, 
where the cool disk is truncated at large radii from the ISCO \citep{DGK07, PV14}.
The source of seed photons for Comptonization is a question intimately related  to the accretion geometry.
In the corona geometry -- where the accretion disk is always thought to extend down to the ISCO -- the disk photons should be the dominant seed photon supplier for Comptonization throughout an X-ray outburst, while in the hot flow geometry the large, variable disk truncation radius (and hot flow size) could make synchrotron radiation from the hot flow itself a more important seed photon source than the disk.
In addition, if the disk is truncated and the truncation radius varies in time, then one expects the reflection amplitude and the spectral hardness to evolve when the number of seed photons changes during the outburst (see, e.g., \citealt{Gilfanov10}).

In this paper we present our study of long-term spectral variations of \swj.
In Sect.~2, we first describe our target and then describe the methods and models we  used for data analysis.
In Sect.~3, we present the results from X-ray spectral analysis of the source using phenomenological models, and in Sect.~4 we focus on self-consistent spectral simulations within the optical to X-ray band.
We discuss the implications of our findings in Sect.~5, in particular how  the broad band energy spectrum can be produced during the late outburst stages by a synchrotron self-Compton (SSC) mechanism in the hot flow, 
rather than by Compton up-scattering soft seed photons from the accretion disk.
We summarize the main results in Sect.~6.

\section{Target and observations}

\begin{figure*}
\centering
\includegraphics[width=17cm,clip]{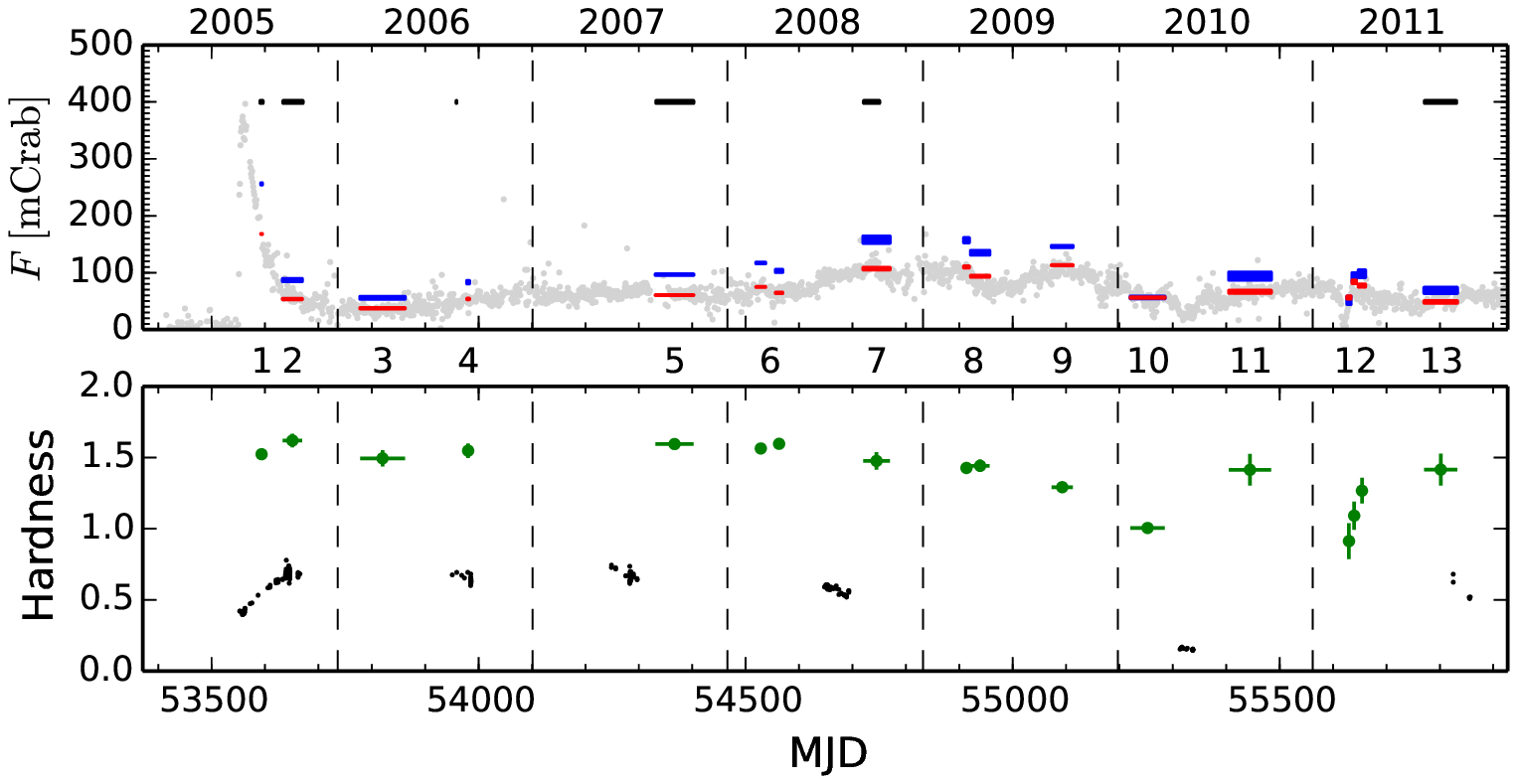}
 \caption{\label{fig:lc} Light curve of the outburst. 
{\it Top panel}: light gray dots show the 15--50~keV \swift/BAT light curve. 
The red and blue stripes show the visibility season averaged \inte/IBIS/ISGRI fluxes for the groups 1--13 in the 20--60~keV and 60--200~keV bands, respectively. 
The horizontal width of the stripes indicate the time interval over which the observations were made, and the vertical width indicate the measured flux error.
The black stripes only indicate the time ranges where PCA spectra were accumulated.
The {\it bottom panel} shows the hardness ratios between the two \inte\ bands (green symbols) as well as the \swift/XRT hardness ratios between the 2$-$10~keV and 0.3$-$2~keV bands (black symbols).
}
\end{figure*}

\subsection{Swift J1753.5--0127}

\swj\ was discovered with the \textit{Swift} Burst Alert Telescope (BAT) on May 30, 2005 \citep{PBC05}.
\swj\ is located at high Galactic latitude and, therefore, the hydrogen column density towards it is relatively low $N_{\rm H}\simeq2\times10^{21}$~cm$^{-2}$ \citep{Cadolle07, FMF14}.
The distance of \swj\ is poorly constrained, but likely lies somewhere between $2.5{-}10$~kpc \citep{Cadolle07, ZDT08}.
Recent optical observations suggest that the binary consists of a low-mass BH and a somewhat evolved $~0.2\,\Msun$ companion star \citep{NVP14}.
We adopt $i=40\degr$ in the present work, since for inclinations higher than ${\sim}60\degr$ the measured mass function would suggest a neutron star primary \citep{NVP14}.

Although many objects show spectral state transitions during the outburst peak, a growing number of BH-LMXBs have been found only in the hard state \citep{BBF04}.
The target of our investigation \swj\ was originally identified as one of these systems, 
although we now know that in 2010 it made a ``failed transition'' towards the soft state (or the hard-intermediate state; \citealt{SMM13}), and since 2011 short incursions into the soft state have been seen \citep{YYN15, SGA16}. 
Unlike a typical LMXB, \swj\ has not returned to quiescence after 10~yr in outburst.
This unique feature has motivated several X-ray spectral and timing investigations (e.g., \citealt{ZQZ07,RS07,DGS08,WU09,RFRM09,RFM10,KvdKU13}).
Particularly relevant previous works with respect to our study are the ones focused on spectral analysis of \swj:
e.g., \citet{CDSG10} explored the spectral behavior of the 2005 outburst peak, the failed transition of 2010 was studied in \citet{SMM13}, and the 2012 transition to soft state was described in \citet{YYN15}.
Our approach was to conduct a comprehensive spectral analysis of the entire data set available to the end of the \xte\ mission, complemented by self-consistent spectral modeling.
For the first time, we focused on the evolution of electron temperatures and the differences in other physical parameters obtained by fitting different spectral models to various stages of the ongoing outburst.

\begin{table*}
\caption{\label{tab:obslog}Observation log.} 
\centering
\tiny
\begin{tabular}{@{}llllllll}
\hline\hline\noalign{\smallskip}
ID      & \multicolumn{2}{l}{\swift\ range or ID and start time}& $T$exp& \multicolumn{2}{l}{\xte\ range or ID and start time}     & $T$exp          & ISGRI range   \\
        & ID            & yy-mm-ddThh:mm                        & [s]   & ID                      & yy-mm-ddThh:mm                & [s]           & yy-mm-ddThh:mm\\
\hline\noalign{\smallskip}
\multicolumn{8}{c}{2005 outburst peak} \\
\hline\noalign{\smallskip}
P00     & 00143778000   & 05-07-01T21:33                        & 999   & 91094-01-01-00  & 05-07-02T01:02                & 1519 / 910    & $...$          \\
P03     & 00030090003   & 05-07-06T17:14                        & 577   & 91094-01-01-04  & 05-07-06T11:31                & 3132 / 1446   & $...$          \\
P04     & 00030090004   & 05-07-06T18:58                        & 1550  & 91423-01-01-04  & 05-07-06T05:13                & 3136 / 1288   & $...$          \\
P07     & 00030090007   & 05-07-08T16:10                        & 884   & 91094-01-02-01          & 05-07-08T17:09                & 2671 / 1670   & $...$   \\ 
P06     & 00030090006   & 05-07-08T17:22                        & 1942  & 91094-01-02-01          & 05-07-08T17:09                & 2671 / 1670   & $...$   \\
P10     & 00030090010   & 05-07-09T08:06                        & 1115  & 91094-01-02-00          & 05-07-09T11:52                & 6305 / 3188   & $...$   \\
P08     & 00030090008   & 05-07-09T09:28                        & 97    & 91094-01-02-00  & 05-07-09T11:52                & 6305 / 3188   & $...$          \\
P09     & 00030090009   & 05-07-09T09:30                        & 1935  & 91094-01-02-00          & 05-07-09T11:52                & 6305 / 3188   & $...$           \\
P11     & 00030090011   & 05-07-10T11:12                        & 106   & 91094-01-02-02          & 05-07-10T11:29                & 3108 / 1523   & $...$           \\
P12     & 00030090012   & 05-07-10T11:14                        & 1937  & 91094-01-02-02  & 05-07-10T11:29                & 3108 / 1523   & $...$          \\
P13     & 00030090013   & 05-07-10T14:40                        & 934   & 91094-01-02-02          & 05-07-10T11:29                & 3108 / 1523   & $...$           \\
P15     & 00030090015   & 05-07-11T08:08                        & 2064  & 91094-01-02-03          & 05-07-11T14:14                & 3156 / 1619   & $...$           \\
P16     & 00030090016   & 05-07-11T10:04                        & 876   & 91094-01-02-03          & 05-07-11T14:14                & 3156 / 1619   & $...$           \\
P18     & 00030090018   & 05-07-21T03:01                        & 175   & 91423-01-03-06  & 05-07-21T21:12                & 3111 / 1691   & $...$          \\
P19     & 00030090019   & 05-07-24T11:26                        & 688   & 91423-01-04-02  & 05-07-24T12:07                & 3130 / 1754   & $...$          \\
P20     & 00030090020   & 05-08-05T00:05                        & 361   & 91423-01-06-00  & 05-08-05T15:07                & 3138 / 1082   & $...$          \\
P21     & 00030090021   & 05-08-23T08:21                        & 836   & 91423-01-08-02  & 05-08-23T11:03                & 1100 / 418    & $...$          \\
P23     & 00030090023   & 05-09-07T12:58                        & 2145  & 91423-01-10-00          & 05-09-06T16:17                & 1930 / 703    & $...$           \\
P24     & 00030090024   & 05-09-11T09:59                        & 2323  & 91423-01-11-00          & 05-09-10T17:47                & 3125 / 1045   & $...$           \\
P26     & 00030090026   & 05-09-25T07:03                        & 2083  & 91423-01-13-00  & 05-09-25T15:21                & 1227 / 520    & $...$          \\
P30     & 00030090030   & 05-10-18T18:56                        & 2586  & 91423-01-16-01  & 05-10-19T18:06                & 1227 / 787    & $...$          \\
P31     & 00030090031   & 05-10-22T08:11                        & 1712  & 91423-01-17-00  & 05-10-22T13:30                & 2062 / 1115   & $...$          \\
\hline\noalign{\smallskip}
\multicolumn{8}{c}{INTEGRAL groups in the outburst tail} \\
\hline\noalign{\smallskip}
IG1     & 00030090020   & 05-08-05T00:05                        & 361   & \multicolumn{2}{l}{05-08-09T02:29 -- 05-08-13T17:31}    & 6585          & 05-08-10T13:33 -- 05-08-12T15:32 \\ 
IG2     & \multicolumn{2}{l}{05-09-19T22:15 -- 05-10-22T08:40}  & 21660 & \multicolumn{2}{l}{05-09-21T05:25 -- 05-10-27T16:46}    & 16170         & 05-09-20T04:53 -- 05-10-26T09:01 \\  
IG3     & $...$         & $...$                                 & $...$ & $...$           & $...$                                 & $...$         & 06-02-12T01:35 -- 06-05-06T23:38 \\ 
IG4     & \multicolumn{2}{l}{06-09-01T15:56 -- 06-09-07T18:16}& 3673    & 92404-01-03-00 & 06-08-11T11:01                 & 687           & 06-08-31T07:13 -- 06-09-04T10:39 \\ 
IG5     & 00030090050   & 07-07-15T18:56                        & 1674  & \multicolumn{2}{l}{07-08-20T11:11 -- 07-10-29T05:41}    & 34060         & 07-08-19T06:59 -- 07-10-29T10:24 \\ 
IG6$_1$ & $...$         & $...$                                 & $...$ & $...$           & $...$                                 & $...$         & 08-02-23T17:29 -- 08-03-12T01:29 \\ 
IG6$_2$ & $...$         & $...$                                 & $...$ & $...$           & $...$                                 & $...$         & 08-03-31T05:03 -- 08-04-13T01:20 \\ 
IG7     & \multicolumn{2}{l}{08-08-15T01:42 -- 08-08-15T04:57}  & 1958  & \multicolumn{2}{l}{08-09-12T09:11 -- 08-10-11T08:41}    & 14840         & 08-09-11T03:39 -- 08-10-30T22:36 \\ 
IG8$_1$ & $...$         & $...$                                 & $...$ & $...$           & $...$                                 & $...$         & 09-03-18T16:53 -- 09-03-28T07:46 \\ 
IG8$_2$ & $...$         & $...$                                 & $...$ & $...$           & $...$                                 & $...$         & 09-03-31T13:24 -- 09-05-05T16:02 \\ 
IG9     & $...$         & $...$                                 & $...$ & $...$           & $...$                                 & $...$         & 09-08-29T23:07 -- 09-10-08T16:50 \\ 
IG10    & $...$         & $...$                                 & $...$ & $...$           & $...$                                 & $...$         & 10-01-24T15:10 -- 10-03-29T20:58 \\ 
IG11    & $...$         & $...$                                 & $...$ & $...$           & $...$                                 & $...$         & 10-07-28T01:27 -- 10-10-14T18:12 \\ 
IG12$_1$& $...$         & $...$                                 & $...$ & $...$           & $...$                                 & $...$         & 11-03-06T13:48 -- 11-03-13T01:57 \\ 
IG12$_2$& $...$         & $...$                                 & $...$ & $...$           & $...$                                 & $...$         & 11-03-15T23:37 -- 11-03-23T02:57 \\ 
IG12$_3$& $...$         & $...$                                 & $...$ & $...$           & $...$                                 & $...$         & 11-03-27T18:40 -- 11-04-09T09:47 \\ 
IG13    & 00031232019   & 11-09-21T00:42                        & 1299  & \multicolumn{2}{l}{11-09-07T17:23 -- 11-09-26T17:02}    & 8485          & 11-07-28T19:22 -- 11-09-28T10:51 \\ 
\hline\noalign{\smallskip}
\multicolumn{8}{c}{Failed transition of 2010} \\
\hline\noalign{\smallskip}
FT08 & 00031232008      & 10-04-27T12:23                        & 1032  &  $...$  & $...$                                 & $...$         & $...$ \\
FT09 & 00031232009      & 10-04-29T03:01                        & 1133  &  $...$  & $...$                                 & $...$         & $...$ \\
FT10 & 00031232010      & 10-05-01T03:13                        & 1179  &  $...$  & $...$                                 & $...$         & $...$ \\
FT11 & 00031232011      & 10-05-03T04:41                        & 1356  &  $...$  & $...$                                 & $...$         & $...$ \\
FT13 & 00031232013      & 10-05-09T03:51                        & 938   &  $...$  & $...$                                 & $...$         & $...$ \\
FT14 & 00031232014      & 10-05-11T03:49                        & 1080  &  $...$  & $...$                                 & $...$         & $...$ \\
FT17 & 00031232017      & 10-05-21T01:58                        & 5030  &  $...$  & $...$                                 & $...$         & $...$ \\
FT18 & 00031232018      & 10-05-21T11:07                        & 1600  &  $...$  & $...$                                 & $...$         & $...$ \\
\hline\noalign{\smallskip}\multicolumn{8}{c}{Soft state of 2012} \\\hline\noalign{\smallskip}
SS20 & 00031232020      & 11-10-20T23:46                        & 1852  &  $...$  & $...$                                 & $...$         & $...$ \\
SS21 & 00031232021      & 11-10-22T19:13                        & 532   &  $...$  & $...$                                 & $...$         & $...$ \\
SS22 & 00031232022      & 12-04-20T07:36                        & 1974  &  $...$  & $...$                                 & $...$         & $...$ \\
SS23 & 00031232023      & 12-04-22T06:07                        & 1476  &  $...$  & $...$                                 & $...$         & $...$ \\
SS24 & 00031232024      & 12-04-24T17:25                        & 1007  &  $...$  & $...$                                 & $...$         & $...$ \\
SS25 & 00031232025      & 12-04-26T09:45                        & 1197  &  $...$  & $...$                                 & $...$         & $...$ \\
SS26 & 00031232026      & 12-04-28T16:21                        & 500   &  $...$  & $...$                                 & $...$         & $...$ \\
SS27 & 00031232027      & 12-04-30T14:45                        & 960   &  $...$  & $...$                                 & $...$         & $...$ \\
SS28 & 00031232028      & 12-05-02T22:57                        & 833   &  $...$  & $...$                                 & $...$         & $...$ \\
SS29 & 00031232029      & 12-05-03T18:07                        & 1117  &  $...$  & $...$                                 & $...$         & $...$ \\
SS30 & 00031232030      & 12-05-06T11:51                        & 990   &  $...$  & $...$                                 & $...$         & $...$ \\
SS31 & 00031232031      & 12-05-08T05:37                        & 1067  &  $...$  & $...$                                 & $...$         & $...$ \\
SS32 & 00031232032      & 12-09-19T10:04                        & 472   &  $...$  & $...$                                 & $...$         & $...$ \\ \hline
\end{tabular}
\tablefoot{{The first column} is the ID used in this work, and during the outburst peak we follow the same data selection 
as in \citet{CDSG10}, Table~1. 
The first letters have the following meaning: P is the outburst peak, IG marks the episodes with \inte\ data, FT denotes failed 2010
transition, and SS describes the soft state (hard X-ray dip) of 2012.
}
\end{table*}

\vspace*{2.5mm}
\subsection{Data analysis}\vspace*{1mm}

We  analyzed archival X-ray spectral data of \swj\  from the onset of the X-ray outburst until the end of \xte\ mission on January 5, 2012 (see Table~\ref{tab:obslog}).
We used data from seven instruments: \inte/IBIS/ISGRI, \swift/UVOT, \swift/XRT, \swift/BAT, \xte/PCA, \xte/HEXTE, and \maxi/GSC.
In the paper we discuss four distinct stages of the outburst that are listed in Table~\ref{tab:obslog}: 
the outburst peak (denoted by letter P), the outburst tail having \inte\ coverage (denoted by IG), the~failed 2010 transition (denoted by FT), and the 
2012 soft state (denoted by SS).

In the hard X-ray domain we used data collected with the IBIS telescope \citep{ULdC03} on board the \inte\ observatory \citep{WCdC03}. 
Spectra and light curves were constructed using the data only from the ISGRI detector layer sensitive to photons with energies from 15~keV to 1 MeV \citep{LLL03}. 
The data were processed with the \inte\ Offline Science Analysis (OSA) version 10.0, provided by the ISDC\footnote{ISDC Data Centre for Astrophysics, \url{http://www.isdc.unige.ch/}}.
We also used \swift/BAT data to illustrate the long-term light curve in the $15$--$50$~keV band, which is shown in Fig.~\ref{fig:lc}.
To obtain the light curve, we used \swift/BAT transient monitor results provided by the \swift/BAT team (see \citealt{KHC13}) and the count rate to flux conversion was done as in \citet{TBM10}.

We analyzed \xte/PCA Standard 2 data using {\sc heasoft} version 6.14. 
The spectral data were extracted from the top layer of PCU 2 in the $3{-}25$~keV energy range. 
Standard methods were used to create spectral responses, to estimate the background spectra, and 
to remove dead-time effects, and the recommended 0.5~percent systematic errors were added \citep{JMR06}.
For the \xte/HEXTE we analyzed the standard spectral products from cluster A in the $25{-}200$~keV range.
We only used HEXTE data that covered the outburst peak in June 2005 because after spring 2006 the cluster stopped rocking. 

\begin{table*}
 \centering
\caption{\label{tab:INTEGRAL}Best fitting parameters from modeling {INTEGRAL} spectra with {\sc powerlaw}, {\sc cutoffpl}, and {\sc nthcomp} models.}
\setlength{\tabcolsep}{5pt}
\renewcommand{\arraystretch}{1.2}
\begin{tabular}{@{}lccccccccccccc}
\hline\hline\noalign{\smallskip}
ID      & $\Gamma$                      & $\chi^2 / {\rm d.o.f.}$& $P_{\rm null}$& $\Gamma_{\rm cpl}$ & $E_{\rm fold}$     & $\chi^2/{\rm d.o.f.}$ & $\Gamma_{\rm Comp}$     & $kT_{\rm e}$ & $\chi^2/{\rm d.o.f.}$ & $P_{\rm Ftest}$         & $F_{\rm \gamma}$      \\
\hline\noalign{\smallskip}
& \multicolumn{3}{c}{{\sc powerlaw}}      & \multicolumn{3}{c}{{\sc cutoffpl}} & \multicolumn{3}{c}{{\sc nthcomp}}       &                       &                       \\
IG$01$          & $1.690_{-0.004}^{+0.004}$     & $267.2/32$    & ${<}1$          & $1.47_{-0.02}^{+0.02}$  & $230_{-20}^{+20}$     & $33.6/31$     & $1.646_{-0.005}^{+0.005}$     & $140_{-20}^{+30}$       & $26.7/31$     & $1.7 \times10^{-15}$  & $6.71_{-0.14}^{+0.11}$ \\
IG$02$          & $1.63_{-0.02}^{+0.02}$        & $38.4/32$     & $20.3$         & $...$                         & $...$                 & $...$         & $...$                           & $...$                 & $...$         & $...$           & $3.23_{-0.10}^{+0.09}$ \\
IG$03$          & $1.64_{-0.03}^{+0.03}$        & $42.7/32$     & $9.8$          & $1.25_{-0.12}^{+0.11}$        & $130_{-30}^{+50}$     & $28.1/31$         & $1.60_{-0.03}^{+0.03}$        & $70_{-20}^{+80}$                      & $32.8/31$       & $3.5 \times10^{-4}$   & $1.33_{-0.2}^{+0.07}$ \\
IG$04$          & $1.62_{-0.02}^{+0.02}$        & $47.6/32$     & $3.7$          & $1.16_{-0.10}^{+0.09}$        & $110_{-20}^{+30}$     & $16.1/31$         & $1.56_{-0.02}^{+0.02}$        & $54_{-9}^{+20}$       & $21.0/31$         & $8.7 \times10^{-9}$   & $1.86_{-0.14}^{+0.07}$ \\
IG$05$          & $1.639_{-0.006}^{+0.006}$     & $172.7/32$    & ${<}1$          & $1.39_{-0.02}^{+0.02}$          & $220_{-20}^{+20}$     & $38.0/31$     & $1.610_{-0.006}^{+0.006}$       & $130_{-20}^{+30}$     & $51.6/31$     & $1.0 \times10^{-11}$    & $2.52_{-0.09}^{+0.06}$\\
IG$06_1$        & $1.674_{-0.007}^{+0.007}$     & $90.8/32$     & ${<}1$          & $1.51_{-0.02}^{+0.02}$          & $320_{-40}^{+50}$     & $29.2/31$     & $1.646_{-0.007}^{+0.007}$       & ${>}280$                & $28.7/31$     & $3.9 \times10^{-9}$     & $3.29_{-0.4}^{+0.05}$ \\
IG$06_2$        & $1.68_{-0.02}^{+0.02}$        & $33.6/32$     & $38.8$         & $...$                         & $...$                 & $...$         & $...$                           & $...$                 & $...$         & $...$           & $3.60_{-0.10}^{+0.08}$ \\
IG$07$          & $1.75_{-0.03}^{+0.03}$        & $32.7/32$     & $43.4$         & $...$                         & $...$                 & $...$         & $...$                           & $...$                 & $...$         & $...$           & $5.5_{-0.3}^{+0.2}$   \\
IG$08_1$        & $1.76_{-0.02}^{+0.02}$        & $41.1/32$     & $13.0$         & $...$                         & $...$                 & $...$         & $...$                           & $...$                 & $...$         & $...$           & $5.51_{-0.2}^{+0.13}$         \\
IG$08_2$        & $1.73_{-0.02}^{+0.02}$        & $30.3/32$     & $55.2$         & $...$                         & $...$                 & $...$         & $...$                           & $...$                 & $...$         & $...$           & $4.80_{-0.14}^{+0.13}$        \\
IG$09$          & $1.837_{-0.007}^{+0.007}$     & $183.1/32$    & ${<}1$          & $1.54_{-0.03}^{+0.03}$          & $171_{-14}^{+20}$     & $39.5/31$     & $1.738_{-0.009}^{+0.010}$       & $130_{-20}^{+40}$     & $55.7/31$     & $7.4 \times10^{-12}$    & $3.71_{-0.12}^{+0.07}$ \\
IG$10$          & $2.02_{-0.02}^{+0.02}$        & $47.8/32$     & $3.6$         & $1.68_{-0.08}^{+0.08}$          & $140_{-30}^{+40}$     & $26.0/31$     & $1.86_{-0.03}^{+0.03}$  & $120_{-40}^{+200}$                    & $28.7/31$         & $1.6 \times10^{-5}$   & $1.47_{-0.2}^{+0.03}$ \\
IG$11$          & $1.71_{-0.07}^{+0.07}$        & $22.0/32$     & $90.8$         & $...$                         & $...$                 & $...$         & $...$                           & $...$                 & $...$         & $...$           & $3.22_{-0.3}^{+0.14}$         \\
IG$12_1$        & $2.16_{-0.09}^{+0.10}$        & $27.5/32$     & $69.2$         & $...$                         & $...$                 & $...$         & $...$                           & $...$                 & $...$         & $...$           & $1.52_{-0.2}^{+0.06}$ \\
IG$12_2$        & $2.03_{-0.07}^{+0.07}$        & $14.3/32$     & $99.7$         & $...$                         & $...$                 & $...$         & $...$                           & $...$                 & $...$         & $...$           & $2.68_{-0.3}^{+0.10}$         \\
IG$12_3$        & $1.90_{-0.06}^{+0.06}$        & $19.9/24$     & $70.2$         & $...$                         & $...$                 & $...$         & $...$                           & $...$                 & $...$         & $...$           & $2.86_{-0.2}^{+0.12}$         \\
IG$13$          & $1.75_{-0.06}^{+0.06}$        & $16.8/32$     & $98.7$         & $...$                         & $...$                 & $...$         & $...$                           & $...$                 & $...$         & $...$           & $2.37_{-0.2}^{+0.11}$         \\
\noalign{\smallskip}\hline 
\end{tabular}
\tablefoot{The first column is the INTEGRAL group ID, $P_{\rm null}$ is the null hypothesis probability (that we can reject the
model), and $P_{\rm Ftest}$ shows the probability of change improvement from the F-test ({{\sc powerlaw}} vs. {{\sc cutoffpl}}).
Flux $F_{\rm \gamma}$ is computed in the $20{-}500$~keV band and the units are in $10^{-9}\,{\rm erg\,cm^{-2}\,s^{-1}}$; the folding energy $E_{\rm fold}$ and the electron temperature $kT_{\rm e}$ are given in~keV.
}\vspace*{-2mm}
\end{table*}

We obtained the \swift/XRT data using the XRT generator tool \citep{Evans2009}.
The XRT spectral data were binned to have at least 20 counts per bin with the {\sc grppha} tool and 3~percent systematic errors were added to the spectral data\footnote{\swift/XRT CALDB release note
SWIFT-XRT-CALDB-09 \href{http://www.swift.ac.uk/analysis/xrt/files/SWIFT-XRT-CALDB-09_v16.pdf}{\tt http://www.swift.ac.uk/analysis/xrt/files/SWIFT-XRT-\\CALDB-09\_v16.pdf}}.
We used the data in the $0.6{-}10$~keV range, except in a few short XRT~\mbox{observations} that had usable data only up to ${\sim}7$~keV.
In \mbox{addition}, some XRT spectra showed saw-tooth-like residuals around the silicon edge at ${\sim}2$~keV, indicative of calibration problems.
In these cases we ignored the energy region around 2~keV. 
We also obtained a \maxi/GSC \citep{MKU09} light curve in the 2$-$4~keV range to highlight the spectral softening episode around the soft state of 2012 (see also \citealt{YYN15}). 
The conversion to Crab rate was done by extracting the mean Crab photon flux in the 2--4~keV range around the same epoch as the \swj\ observations.

We also used \textit{Swift}/UVOT data. 
These observations were processed and analyzed using the \textit{Swift} release 3.7 software
within the {\sc heasoft} suite, together with the most recent version of the Calibration Database. 
The data were reduced following the procedure described in \citet{PBP08}. 
The UVOT fluxes were calculated using a standard 5$''$ extraction region;  the background was estimated from four nearby circular source-free \mbox{regions}.

Our aim was to measure the broad band $0.6{-}200$~keV spectra with \swift/XRT, \xte/PCA/HEXTE, and \inte/IBIS/ISGRI.
During the initial outburst peak of 2005, we reanalyzed the same \swift/XRT and \xte/PCA/HEXTE data sets as \citet{CDSG10} to facilitate comparisons between our results (we call these groups P00 to P31; see Table~\ref{tab:obslog}).
In the latter part of outburst we defined 13 \inte\ groups (IG01 to IG13, see Table~\ref{tab:obslog} and Fig.~\ref{fig:lc}) because \inte\ can only observe \swj\ during two visibility windows in the spring and autumn.
Within each \inte\ group (typical \mbox{exposure} times are a few hundred ks) we analyzed all \mbox{simultaneous} \xte/PCA data, except for IG04, where only one quasi-simultaneous \xte/PCA observation was available.
If significant variability was detected in the \xte/PCA or the \inte/IBIS/ISGRI data, we divided the groups into subgroups. 
We then analyzed each simultaneous (or \mbox{quasi-simultaneous}) \swift/XRT spectrum close to these groups, 
and we included these data in our analysis only if the spectral index was similar to \xte/PCA spectra in the common $3{-}10$~keV range.
We also analyzed the \swift/XRT data that were taken during the failed transition of 2010 (FT08 to FT18) and during the 2012 soft state transition (SS20 to SS32).
The data sets that were used in the analysis are listed in Table~\ref{tab:obslog}, together with the best fitting parameters (Tables~\ref{tab:INTEGRAL},~\ref{tab:nthComp}, and~\ref{tab:COMPPS}) that were obtained with {\sc xspec} version 12.8.1.
We quote 68~percent confidence intervals as parameter errors throughout the paper.

\subsection{Spectral models}

We used a combination of several simple phenomenological spectral models to study the outburst behavior.
For the \inte\ groups, we modeled the data with a {{\sc powerlaw}} (flux is $F(E) \propto E^{-(\Gamma-1)}$, where $\Gamma$ is the photon index), 
or a {{\sc cutoffpl}} ($F(E) \propto E^{-(\Gamma-1)}\exp(-E/E_{\rm fold})$ model, where $E_{\rm fold}$ is the e-folding energy) or the {\sc nthcomp} model \citep{ZJM96, ZDS99}.
The {\sc nthcomp} model is described by several parameters: 
$\Gamma$, the electron temperature $T_{\rm e}$, and the seed photon temperature $T_{\rm seed}$, which we assumed to be disk-blackbody-like ({\sc diskbb}; \citealt{M84}) and is characterized by the inner disk temperature $T_{\rm dbb}$.
The {{\sc cutoffpl}} and {\sc nthcomp} models were used  only when the simpler {{\sc powerlaw}} model could be rejected with 95~percent \mbox{confidence}. 
To test when the spectral cutoff was statistically significant we used the {\sc ftest} routine in {\sc xspec} (see Table~\ref{tab:INTEGRAL}).

When analyzing the joint, \xte/PCA/HEXTE, \swift/XRT, and \inte/ISGRI spectra we also used the {\sc compps} model assuming spherical geometry \citep{PS96}.
The relevant parameters are $T_{\rm seed}$, $T_{\rm e}$, and the Compton parameter $y \equiv 4 kT_{\rm e}\, {\rm max}[\tau, \tau^2] / 511\, [{\rm~keV}]$ (here $\tau$ is the optical depth of the hot medium).
In addition, we used the {\sc reflect} \mbox{convolution} model \citep{MZ95} for Compton reflection off the accretion disk;
the parameters are the reflection scaling factor $\Omega / 2\pi$ ($\Omega / 2\pi = 1$ for an isotropic source above the accretion disk) and inclination $i$.
Both the {{\sc diskbb}} and {{\sc compps}} model normalizations, $N_{\rm dbb}$ and $N_{\rm c}$, have the same form $(R_{\rm in} [{\rm km}] / D_{10~\rm kpc})^2 \cos i$, where $D_{10~\rm kpc}$ is the distance in units of 10~kpc.
We also use a {\sc constant} parameter to address  cross calibration uncertainties between the instruments and to take into account the fact that the data were obtained quasi-simultaneously.
We fix this parameter to unity for \xte/PCA, and let \swift/XRT and \inte/ISGRI constants vary.
The {\sc wabs} model is used for interstellar absorption with a fixed hydrogen column density $N_{\rm H} = 2\times10^{21}$~cm$^{-2}$ \citep{Cadolle07, CDSG10, FMF14}.
Solar abundances are assumed for absorption and for reflection from the disk.

\begin{figure*}
\centering
\includegraphics[width=15cm,clip]{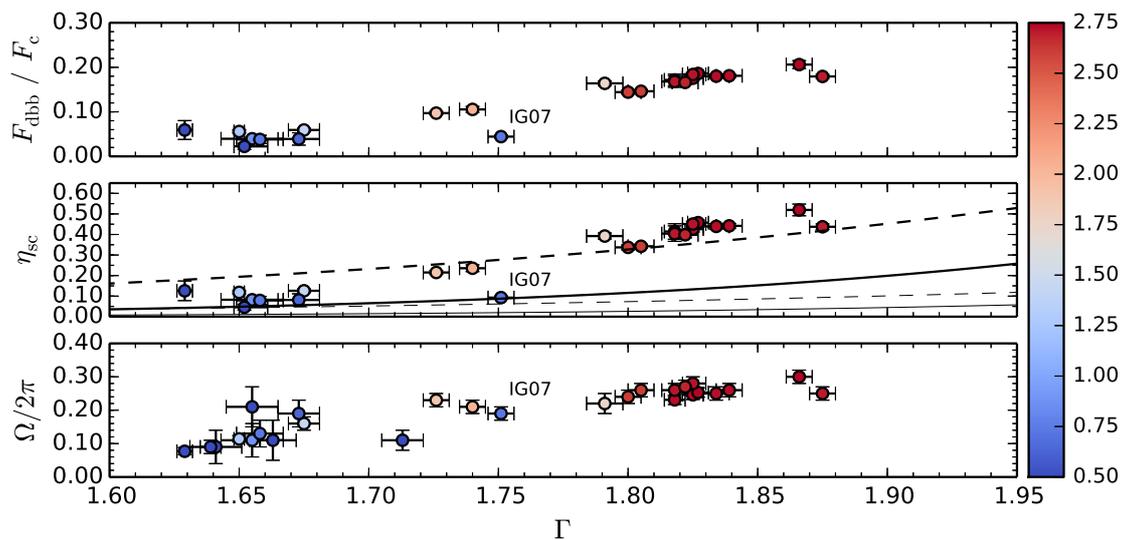}
 \caption{\label{fig:parameters}Evolution of spectral parameters as a function of the photon index $\Gamma$ from the {{\sc nthcomp}} fits. 
Color-coding indicates the estimated bolometric flux in units of $10^{-8}\,\ergcms$. 
The {\it top panel} shows the ratio $F_{\rm dbb}/ F_{\rm c}$ and the {\it middle panel} shows the ratio $F_{\rm seed}/ F_{\rm cor}$, which is connected to the photon index $\Gamma$ of Comptonization power law according to Eq.~(\ref{eq:ratio_theor}).
The data are consistent with the $\delta = 1/6$ trend line (dashed lines) when the flux is above a critical limit $F_{\rm x} \gtrsim 10^{-8}\,\ergcms$ and the spectra is softer than $\Gamma \gtrsim 1.8$. 
When the flux drops below this limit ($F_{\rm x} \lesssim 10^{-8}\,\ergcms$) and the spectrum hardens to $\Gamma \lesssim 1.7$, the data are instead consistent with the $\delta = 1/10$ track (solid lines).
Thick lines correspond to the assumption $i=40\degr$ and thin lines correspond to $i=80\degr$.
This behavior suggests that during the initial outburst peak the seed photons for Comptonization come predominantly from the accretion 
disk, 
whereas below the critical limit -- when the accretion disk has receded further from the BH -- the seed photon come predominantly from synchrotron emission from the inner hot flow.
The {\it bottom panel} shows that the reflection amplitude $\Omega/2\pi$ decreases during the flux decay, indicating that the reflector (i.e., the cold accretion disk) truncates farther from the BH as the outburst progresses.}
\vspace*{4mm}\end{figure*}

\begin{figure*}
\centering
\includegraphics[width=15cm,clip]{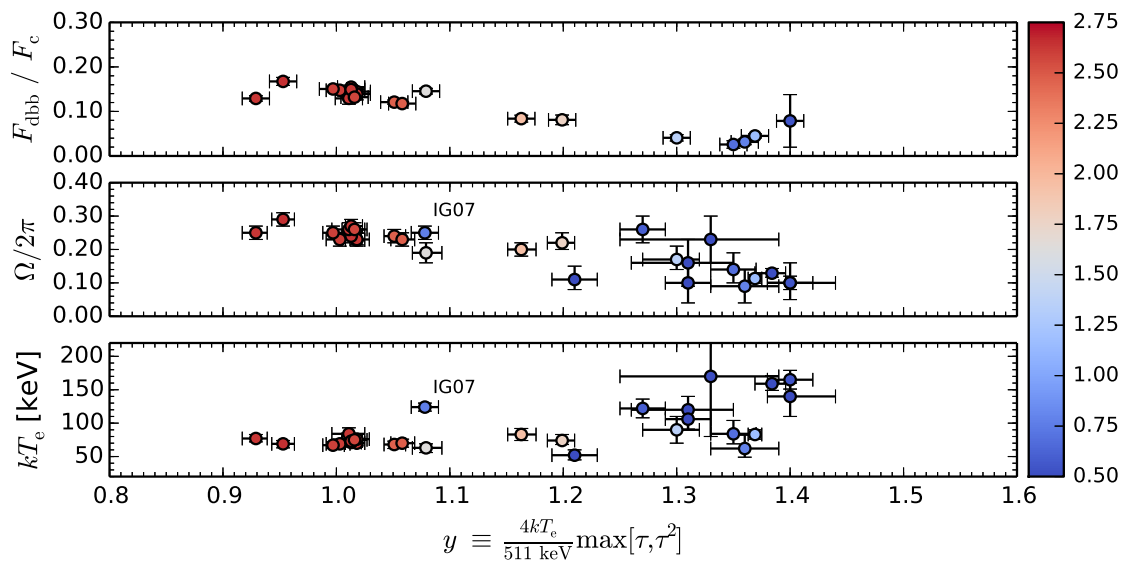}
 \caption{\label{fig:parametersCOMPPS}Evolution of spectral parameters as a function of the Compton parameter $y$ from the {{\sc compps}} fits (color-coding as in Fig.~2).
The overall trends of $L_{\rm dbb}/ L_{\rm c}$ ratio ({\it top panel}) and the reflection amplitude $\Omega/2\pi$ ({\it middle panel}) show qualitatively similar behavior to the {\sc nthcomp} fits.   
We also see how the electron temperature $kT_{\rm e}$ ({\it bottom panel}) increases when the flux drops below the critical limit of $F_{\rm x} \lesssim 10^{-8}\,\ergcms$.
The $kT_{\rm e}$ increase may be caused by the change of geometry and seed photon source: disk truncation leads to fewer disk photons entering the Comptonized region, resulting in less efficient cooling of the electrons and thus higher equilibrium temperatures.
}
\end{figure*}

\section{Spectral variability of \swj\ during the outburst}

\subsection{Long-term outburst evolution seen by INTEGRAL/ISGRI} \label{sec:dips}

The hard X-ray light curve of the outburst is characterized by the pronounced ${\sim}400$ mCrab peak in 2005, followed by the flux decay to ${\sim}50$~mCrab, and significant spectral hardening by the end of that year (IG02).
From 2006  to spring 2008 (IG03$-$IG06$_1$), \swj\ started a gradual  monotonic brightening with a constant spectral hardness and shape (see Fig.~\ref{fig:lc} and 
Table~\ref{tab:INTEGRAL}). 
From spring 2008 to early spring 2009 (IG06$_2$--IG08$_1$) a marked increase in the hard X-ray flux is seen, which is accompanied by clear spectral changes.
The best fitting parameters indicate that the spectra start to soften and, at the same time, the high-energy cutoff was no longer detectable.
After the \mbox{secondary} outburst peak in spring 2009, the hard X-ray flux decline is more complex with several secondary minima and \mbox{maxima}. 
These shorter term fluctuations are superimposed to the ${\sim}420$~day super-orbital periodicity, which starts to become more pronounced in the \swift/BAT data \citep{SCB13}.
In consequence, only strictly simultaneous \swift/XRT data could be used thereafter (from IG08 onward).

From IG09 onward spectral softening becomes more pronounced in the \inte/IBIS/ISGRI and the \swift/XRT hardness ratios shown in the bottom panel of Fig.~\ref{fig:lc}.
Furthermore, the hard X-ray spectra  start displaying the cutoff again (IG09), culminating in a nearly flat, $\Gamma \simeq 2$ spectrum in spring 2010 (IG10).
The softening in hard X-rays during IG10 preceded the failed spectral- and timing transition of summer 2010 \citep{SMM13}, during which the \swift/XRT spectra were softer than in the outburst peak.
After the failed transition was over, the spectrum hardened significantly and the cutoff disappeared again (IG11--IG13).

The spectra after 2010 (IG11--IG13) are similar to those observed in 2006--2008 (IG03--IG06$_1$); hardness is comparable and the spectra show no signs of the high-energy cutoff.
However, the most marked difference is the appearance of hard X-ray dips in the \swift/BAT light curve.
The first  was captured by \inte/IBIS/ISGRI (IG12; see Fig.~\ref{fig:lc}), which detected a gradual spectral hardening back to the ``pre-dip'' level.
Our analysis confirms the conclusions by \citet{SMM13} who studied the failed transition of 2010 and by \citet{YYN15} who analyzed the 2012 dip: the hard X-ray dips are always related to softening of the energy spectrum and they are caused by spectral transitions to (or towards) the soft state.

\begin{figure}
\centering
\includegraphics[width=8.5cm,clip]{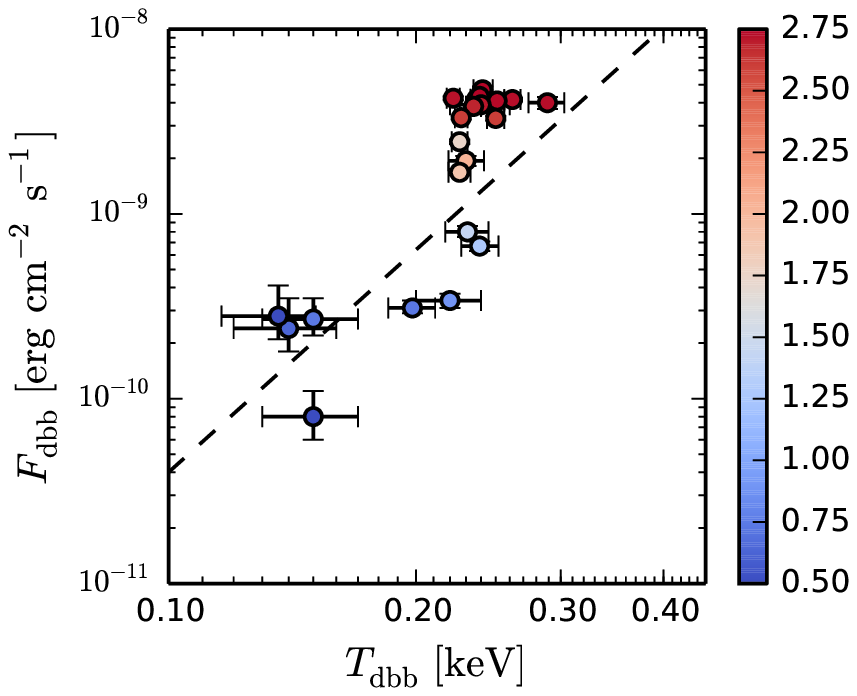}
\includegraphics[width=8.5cm,clip]{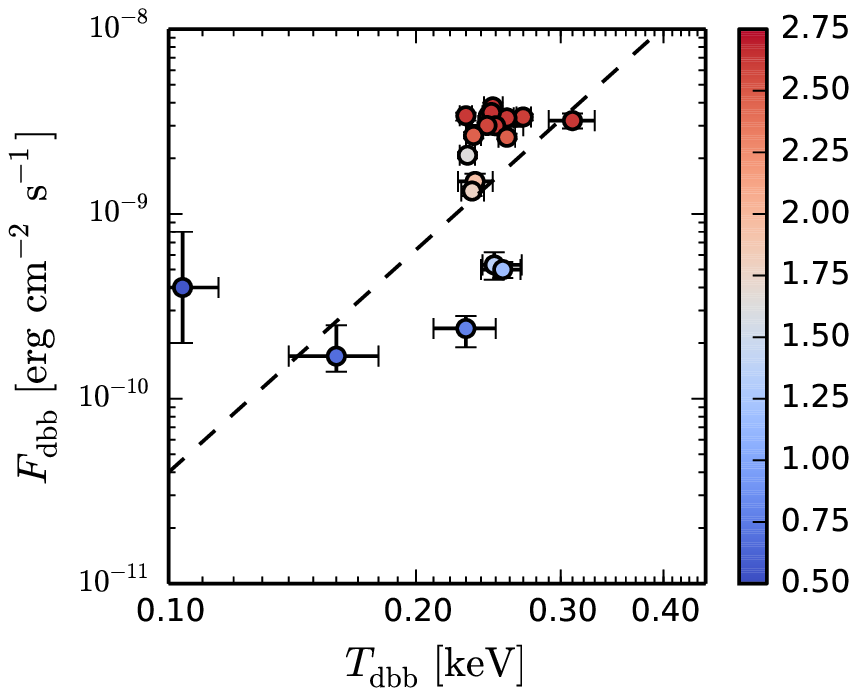}
 \caption{\label{fig:FT}Flux-temperature relation for the disk component from the {\sc nthcomp} fits ({\it top}) and {{\sc compps}} fits ({\it bottom}; color-coding as in Fig.~2).
The dashed line shows a $F \propto T^4$ track to guide the eye.
}\vspace*{4mm}
\end{figure}

\subsection{Comparison between the peak and the tail}

To understand the spectral behavior during the entire X-ray outburst, we decided to reanalyze the spectral variations during the outburst peak.
In our analysis, we used two simple phenomenological models: 
(1) {{\sc constant}~$\times$ {\sc wabs}~$\times$ ({\sc diskbb}~$+$ {\sc relf}~$\ast$ {\sc nthcomp})} (we find very similar best fitting parameters to \citealt{CDSG10}, apart from small differences towards the end of the initial outburst peak; see Table~\ref{tab:nthComp}); and
(2) {{\sc constant}~$\times$ {\sc wabs}~$\times$ ({\sc diskbb}~$+$ {\sc relf}~$\ast$ {\sc compps})}, see Table~\ref{tab:COMPPS}.
The {{\sc diskbb}} model component was added in cases where a simpler model could be rejected with 95~percent \mbox{confidence}.
The {{\sc diskbb}} \mbox{temperature} was fixed to the seed photon temperature~$T_{\rm seed}$. 
In a few cases where the {{\sc diskbb}} component was not significant, $T_{\rm seed}$ was also unconstrained, and in those cases it was fixed to $T_{\rm seed} = 5$~eV.

The most noticeable evolution in the outburst peak is the hardening of spectra as the flux decays, 
which is shown by color-coding the parameter evolution in Figs.~\ref{fig:parameters} and~\ref{fig:parametersCOMPPS} for the {\sc nthcomp} and {{\sc compps}} fits, respectively.
Like \mbox{\citet{CDSG10}}, we also observe the simultaneous decrease of the reflection amplitude from $\Omega/2\pi \simeq 0.25$ to $\simeq\!0.15$ during the decay, which can be interpreted as a signature of the cold disk receding from the BH \citep{ZLS99, GCR99, RGC01, Gilfanov10}.
The relatively high reflection amplitude is accompanied by a pronounced soft excess that also suggests the presence of a cool, optically thick accretion disk close to the BH during the outburst peak.
However, as noted by \citet{CDSG10}, the evolution of the {{\sc diskbb}} model parameters is peculiar.
The inferred inner disk radius $R_{\rm in}$ derived from the {{\sc diskbb}}~model normalization $R_{\rm in}$ decreases towards the end of the initial 2005 outburst peak, 
indicating that the truncation radius decreases as the outburst progresses.
This behavior is in broad disagreement with the predictions of the truncated disk scenario, 
which seems to  produce the dependencies with the spectral slope, reflection amplitude, and variations in the quasi-periodic signals in the outburst peak \citep{CDSG10}.
Our tests and the {{\sc compps}} analysis shows that the observed behavior is largely model independent (compare upper and lower panels of Fig.~\ref{fig:FT}):
while the disk flux decays by more than an order of magnitude we see a decrease in the apparent inner disk radius, but we do not see any change in the inner disk temperature of $T_{\rm dbb} \simeq 0.25$~keV in the six measurements taken during this decay phase.

\begin{figure}
\centering
\includegraphics[width=8.5cm,clip]{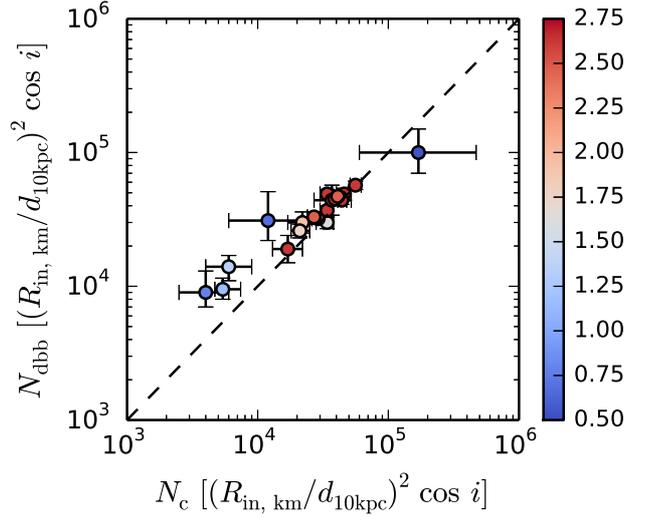}
 \caption{\label{fig:NdiskNcomp} Relation between {{\sc diskbb}} model normalization and the {{\sc compps}} model normalization (color-coding as in Fig. 2).
The dashed line shows a linear, one-to-one trend that may be a sign that about  half of the disk photons enter the hot flow (or the corona above the accretion disk).
The trend seems to disappear when the flux drops below the critical limit.
}\vspace*{4mm}
\end{figure}

In our {{\sc compps}} analysis we detected a new feature. 
We find a striking similarity between the {{\sc diskbb}}- and {{\sc compps}} model normalizations. 
In the outburst peak -- when the flux is above the critical limit --  they attain exactly the same values (see Fig.~\ref{fig:NdiskNcomp}).
Only when the flux drops below the critical limit the one-to-one relation is broken.
Because the {{\sc diskbb}} and {{\sc compps}} model normalizations have the same form $N_{\rm dbb} = N_{\rm c} = (R_{\rm in} [{\rm km}] / D_{10\rm kpc})^2 \cos i$ and because the {{\sc compps}} seed photon temperature was tied to the {{\sc diskbb}} inner disk temperature, i.e., $T_{\rm seed} = T_{\rm dbb}$, the seed photon flux for Comptonization is identical to the {{\sc diskbb}} model soft excess flux. 
The simplest interpretation of this finding is that the Comptonized region intercepts only about 50~percent of the disk photons during the
outburst peak.

For a Comptonized spectrum the ratio between the seed photon and Comptonized power $\eta_{\rm sc}$
can be connected to the photon index $\Gamma$ (see \citealt{B99ASP}) as $\displaystyle \Gamma=\frac{7}{3}\eta_{\rm sc}^\delta$, where $\delta\approx1/6$ for typical temperatures of disk seed photons in
Galactic X-ray binaries $T_{\rm seed}\sim0.2$~keV and $\delta\approx1/10$ for $T_{\rm seed}\sim5$~eV.
In our notations, the formula reads as
\begin{equation}\label{eq:ratio_theor}
\Gamma  \approx \frac{7}{3} \left[F_{\rm dbb}/ \mu (F_{\rm c}-F_{\rm dbb})\right]^{\delta},
\end{equation}
where factor $\mu \equiv \cos i$ comes from the relation between the observed and seed disk photon fluxes.
When comparing these theoretical predictions with the observations in the middle panel of Fig.~\ref{fig:parameters}, 
we find that initially \swj\ follows the $\delta\approx1/6$, $T_{\rm seed}\sim0.2$~keV track (dashed line),
but as soon as the flux drops below a critical limit of $F\sim\!1.5\times 10^{-8}\,\ergcms$ the data are more consistent with 
the $\delta\approx1/10$, $T_{\rm seed}\sim5$ eV track (solid line).

\begin{figure*}
\centering
\includegraphics[width=15cm,clip]{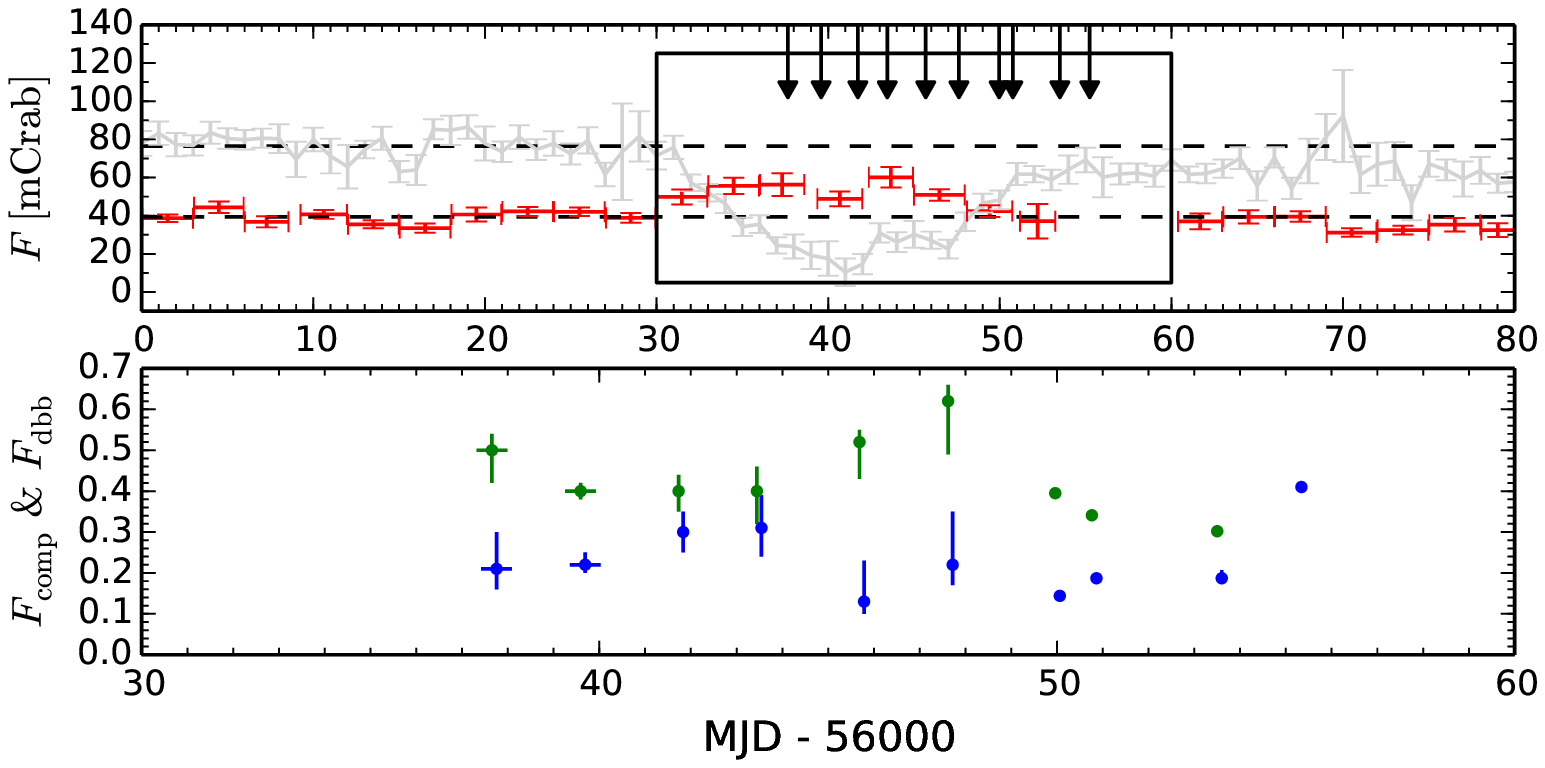}
 \caption{\label{fig:2012dip}Light curves of the 2012 soft state transition (see also \citealt{YYN15}). 
{\it Top panel}: 15--50~keV \swift/BAT (light gray) and 2--4~keV \maxi/GSC (red) light curves anti-correlate, indicative of spectral softening as seen by \inte/ISGRI in the 2011 hard X-ray dip. The arrows indicate the times of the \swift/XRT observations.
{\it Bottom panel}: evolution of the disk (green) and  Comptonized fluxes (blue) from \swift/XRT~{\sc compps} fits. We note how, for the only time during the ten-year outburst of \swj, the {{\sc diskbb}} component emits most of the radiative flux.
}
\end{figure*}

The results from the {{\sc compps}} model fits also indicate that there are two causes for the hardening during the outburst.
In the bottom panel of Fig.~\ref{fig:parametersCOMPPS} we show the relation between the Compton $y$-parameter and the electron temperature $T_{\rm e}$.
We see that during the initial, bright outburst phase, the spectral hardening happens at a constant $T_{\rm e} \simeq 70$~keV and $y \simeq 0.93$--$1.06$.
Because the Compton parameter is related to the optical depth through $y = 4 T_{\rm e} {\rm max}[\tau, \tau^2] / 511\, [{\rm~keV}]$, this hardening arises because the optical depth of the plasma increases from $\tau \simeq 1.3$ to $\tau \simeq 1.4$.
However, right when the $F_{\rm dbb}/F_{\rm c}$ ratio drops to the $\delta = 1/10$~track, 
we observe a twofold increase in electron temperature to $T_{\rm e} \sim 140$~keV, whereas the smaller increase of $y \simeq 1.35$ shows that optical depth actually drops to $\tau \sim 1.1$ in this transition.

The \inte\ data after 2006 brought valuable new information.
For example, the data from IG01 to IG06 helped us to confirm that electron temperature is really as high as $T_{\rm e} \sim 140$~keV.
Furthermore, the IG07 group in particular is worth highlighting as it stands out in Figs.~\ref{fig:parameters} and~\ref{fig:parametersCOMPPS}.
In the {\sc nthcomp} fits, the IG07 $F_{\rm dbb}/F_{\rm c}$ ratio is clearly in the $\delta = 1/10$ track, 
suggesting that from  late 2005 up to at least late 2008 the disk was in a similar state.

\subsection{Transitions to the soft state}

\swj\ has shown clear spectral state transitions since the 2010; 
the failed transition in summer 2010 that was reported in \citet{SMM13} and recently \citet{YYN15} also published their analysis of the transitions seen from 2011 onward using \maxi\ and \swift/XRT data.
Our analysis broadly confirms these published findings about the nature of these transitions, but our comparison of these transitions to the whole outburst allows us to draw some additional conclusions.

We find that there is a clear difference in the strength of the disk component between the spectra during these short incursions towards the soft state and the rest of the outburst.
For the first time the disk is the most dominant spectral component (see Fig.~\ref{fig:2012dip}).
However, the results are strongly model dependent, and the lack of simultaneous hard X-ray data make the comparisons challenging.
In the {\sc nthcomp} fits, we see that the spectra in some observations are not improved by having the additional disk component (such as FT11 or SS24).
Furthermore, the best fitting parameters for $\Gamma$ are highly atypical for a BH; photon \mbox{indices} as high as $\Gamma \sim 4$ have never been observed in BHBs with instruments that are sensitive to photon energies above ${\sim}10$~keV.

The {{\sc compps}} model results were more consistent among themselves, especially after fixing the electron temperature to a typical $T_{\rm e}=100$~keV value.
However, the $y$-parameter indicates that the optical depth is $\tau \sim 0.2$, which is again an atypical value for a thermal Comptonization spectrum of a BH-LMXB.
In order to have $\tau \sim 1$ the electron temperature should be as low as $T_{\rm e} \sim 20$~keV, which is inconsistent with the detection of emission up to ${\sim}200$~keV in the 2011 state transition (IG12$_1$).
This suggests that the hard X-ray tail in the soft state spectra is more likely produced by Comptonization by a non-thermal 
or hybrid electron distribution \citep{McConnell02,PV09,MB09}, but in the absence of simultaneous \swift/XRT and \inte/ISGRI data this 
speculation cannot be confirmed.

\begin{figure}
\centering
\includegraphics[width=8.8cm,clip]{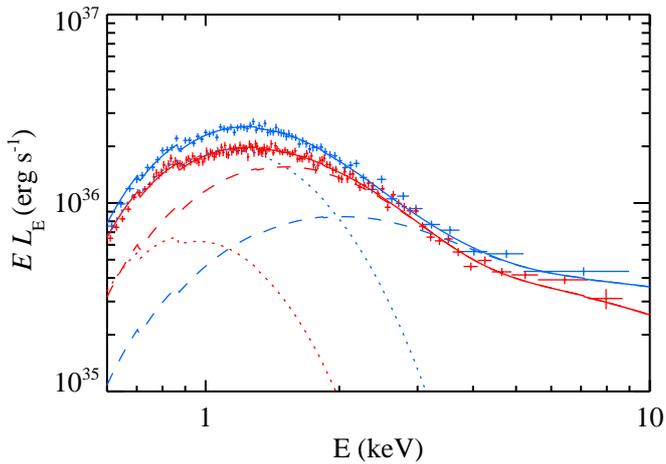}
 \caption{\label{fig:spec2010_2012}\swift/XRT X-ray spectra of \swj\ observed in 2010 (ID 00031232013, red) and 2012 (ID 00031232027, blue). 
The {{\sc diskbb}} model components are shown by dotted lines and the {{\sc compps}}~model components by dashed lines for 2010 and 2012, respectively (distance of 3 kpc is assumed).}
\end{figure}

In the {{\sc compps}} fits  two thermal disk-like components and a high energy tail are clearly needed, which is why we did not fix the inner disk temperature to the seed photon temperature for Comptonization.
The best fitting parameters indicate that the spectra in the 2010 failed transition and the 2012 soft state are somewhat different from each other 
(see Fig.~\ref{fig:spec2010_2012}, and Tables~A.1 and A.2).
The broader shape of thermal component is likely caused by lower disk temperatures and larger truncation radii in the 2010 failed transition than in  the 2012 soft state.
Interestingly, we find that in the 2010 failed transition the normalizations are very similar to the values found at the onset of the 2005 outburst, whereas in the 2012 soft state the {{\sc diskbb}} normalizations are similar to the values found at the end of the main 2005 outburst peak.
While it is widely accepted that in the soft state the thin disk indeed reaches the ISCO, it is much more debated in the hard state (see, e.g., \citealt{DGK07} for discussion).
Our results indicate that in \swj\ the soft excess has a rather stable size scale, seen both towards the end of the 2005 hard state outburst peak and during the soft state incursions.
To get an order of magnitude estimate of the ``corrected'' inner disk radius,  a color correction $f_{\rm c}$ and inner disk \mbox{boundary} correction $\xi$ must be applied to the value obtained from the {{\sc diskbb}} model \mbox{normalization}.
The canonical values are $f_{\rm c} = 1.7$ and $\xi=0.412$ and the corrected radius is obtained through $R_{\rm in,c} = R_{\rm in} \xi f_{\rm c}^2$ (see, e.g., \citealt{KTM98}).
If we take the mean value of the {{\sc diskbb}} model normalization in the 2012 soft state, $N_{\rm dbb} \simeq 19\,000$, and reasonable values $i=40$, $d_{10 \rm kpc}=0.25$, we get a corrected inner disk radius estimate of $R_{\rm in,c} \sim 47$~km (about $5.3R_{\rm S}$ for a $3~\Msun$ BH).
Although the inner disk may be equally likely be at $R_{\rm in, c}\sim 200$~km (if the distance is instead 8~kpc), we cannot exclude the possibility that the inner disk reaches the ISCO.

\section{Self-consistent physical modeling}\label{sect:model}

The phenomenological modeling with the disk Comptonization is in good agreement with the data at the beginning of the outburst.
However, in order to explain the high Comptonization-to-disk ratio observed after P20 (August~5, 2005), according to Eq.~(1) a binary inclination of $i\ga80^\circ$ is required (see the thin dashed line in Fig.~\ref{fig:parameters}, and  discussion in Sect.~5), which is in conflict with the absence of X-ray dips/eclipses and dynamical mass constraints \citep{NVP14}.
Alternatively, the Comptonization continuum can be produced by up-scattering of synchrotron photons that are produced internally in the hot medium.
In this case the spectra resemble power laws down to the energies of the synchrotron seed photons, typically in the optical domain.
The \swift/UVOT data from the outburst tail were indeed shown to be lying on the continuation of the X-ray power law by \citet{CDSG10}.

For reasonable magnetic fields of the hot flow (in or below the equipartition with the ion energy density) and for the electron temperatures and Thomson optical depths found by our phenomenological analysis, most of the synchrotron photons produced by the thermal electron population are internally self-absorbed (see \citealt{WZ00}).
However, the continuous heating and acceleration processes in the hot flow leads to the development of weak non-thermal tails complementing the mostly thermal electron distribution.
These relativistic, non-thermal electrons are major producers of synchrotron seed photons for Comptonization, and they also emit above 100~keV photon energies via inverse Compton scattering \citep[see, e.g.,][]{WZ01}: 
such non-thermal $\gamma$-ray tails are seen in the spectra of Cyg X-1 \citep{McConnell02} and GX 339--4 \citep{DBM10}.

Self-consistent modeling of non-thermal relativistic plasmas is a challenging task because the shape of the entire electron distribution depends on 
the heating and cooling processes rather than the Maxwellian temperature determined by a heating/cooling balance.
Therefore, the problem must be treated numerically \citep{VP09, MB09, VPV13}.
A comparison of modeled spectra obtained from self-consistent electron distributions against the observed spectra 
has been performed for the BH binaries Cyg~X--1 and GX~339--4 in the X-ray/$\gamma$-ray regime \citep{PV09, MB09, DBM10, dSMB2013}.
However, simultaneous comparison with the optical/UV data has not been presented to date.
In this section we perform such self-consistent spectral modeling for the characteristic spectra of \swj\ from optical to X-rays.
We aim to investigate whether the aforementioned difficulties of the disk Comptonization scenario arise because in the outburst tail the spectra are in reality produced with the synchrotron self-Compton mechanism.

\subsection{Model setup}\label{sect:model_spec}

We approximate the emitting hot medium with a homogeneous and isotropic sphere of radius $R$, which is filled with electron gas of Thomson optical depth $\tau$.
The radiation field is characterized by the emitted luminosity $L$ and the medium carries a magnetic field $B$, which we assume to be tangled.

The energy is given to the electrons in the form of a power-law distributed electron injection $n_{\rm e}(\gamma)\propto\gamma^{-\Gamma_{\rm inj}}$, where $n_{\rm e}$ is the electron number density per unit Lorentz factor $\gamma$.
They subsequently cool and thermalize by multiple radiative (\mbox{synchrotron}, Compton, bremsstrahlung) and pair processes 
and interact via Coulomb collisions.
To calculate the radiation spectrum, we solve coupled electron and photon kinetic equations using the code described 
in \citet{VP09}, that was later on extended to include bremsstrahlung processes by \citet{VVP11}.
Equations are solved until self-consistent steady-state solutions for electron and photon distributions are achieved.
We note that typical cooling timescales are much shorter than the viscous timescale (for relevant scalings, see \citealt{VPV13}).

\begin{figure*}
\centering
\begin{tabular}{cc}
 \includegraphics[width=8cm,clip]{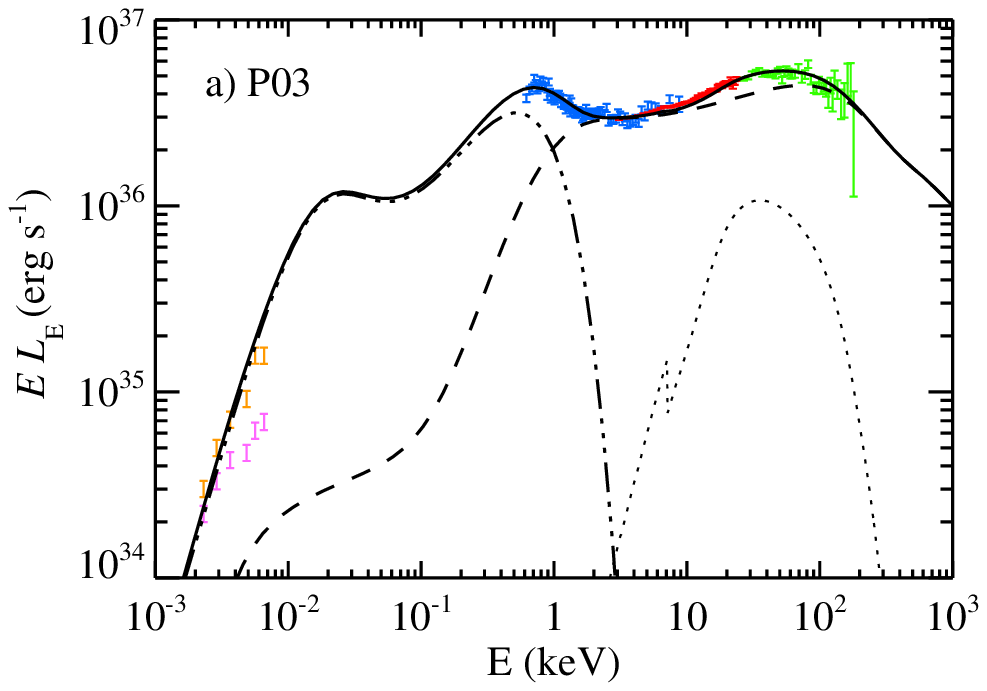} &   \includegraphics[width=8cm,clip]{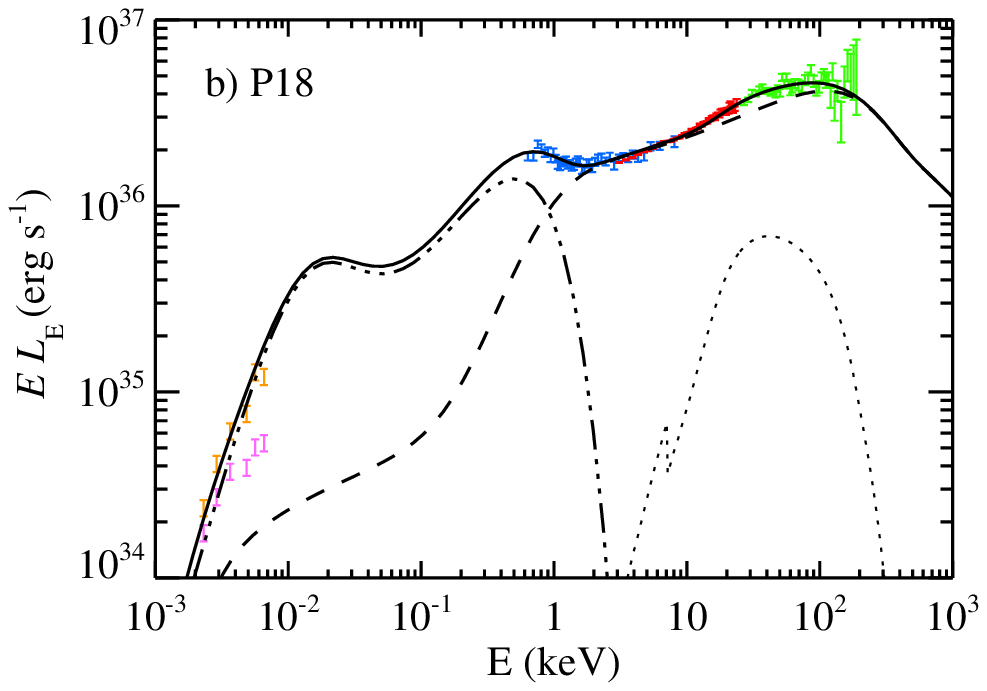}\\
  \includegraphics[width=8cm,clip]{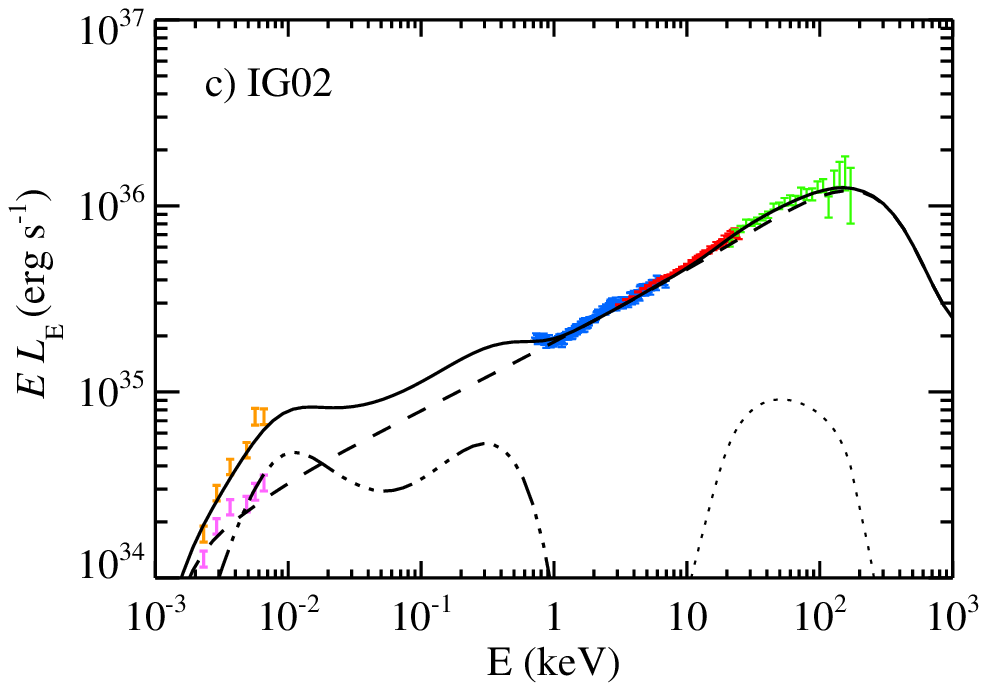}& \includegraphics[width=8cm,clip]{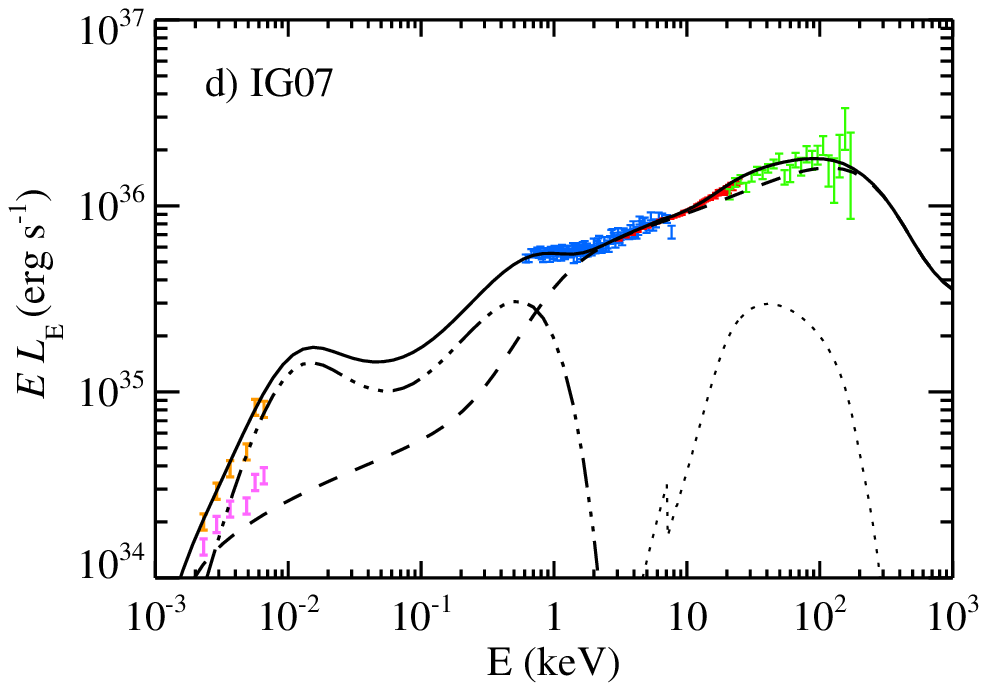} 
\end{tabular}
\caption{\label{fig:simulations} 
Spectra obtained from self-consistent simulations for epochs P03 {\bf a)}; P18 {\bf  b)}; IG02 {\bf c)}; and IG07 {\bf d)}.
The X-ray data from \swift/XRT, \xte/PCA, and \xte/HEXTE (panels {\bf a)} and {\bf b)}) or \inte/ISGRI (panels {\bf c)} and {\bf d)}) are shown with blue, red, and green symbols, respectively.
{\it Swift}/UVOT data, corrected for interstellar reddening, are plotted with magenta symbols for $E(B-V)=0.34$ and with yellow symbols for $E(B-V)=0.45$.
The solid black line is the spectrum obtained from self-consistent simulations and it consists of three major components:
(1) the emission from the hot medium produced by synchrotron radiation and Comptonization of disk and 
synchrotron seed photons with an additional contribution from bremsstrahlung emission (dashed lines);
(2) the contribution of irradiated accretion disk (dot-dashed lines); and
(3) the Compton reflection hump taken from the best fitting {sc compps} model fit (dotted lines).}
\end{figure*}

\begin{table*}
\caption{\label{tab_simulat}
Parameters of the spectral simulations.} 
\centering
     \begin{tabular}{@{} rccccccc@{}}
\hline\hline\noalign{\smallskip}
ID/Parameter      & $R$~($\times10^7$~cm)  & $\tau$  & $L_{\rm hot}$~($\times10^{36}$~erg~s$^{-1}$) &  $B$~($\times10^{5}$~G) 
                                &  $L_{\rm d}$~($\times10^{36}$~erg~s$^{-1}$)   & $T_{\rm dbb}$~(keV) & $T_{\rm irr, out}$~(eV)  \\
\hline\noalign{\smallskip}
 P03                        & 1.0           &  2.0      &  25              & 8    & 5.0   &  0.23  &  3.53 \\
P18                        & 1.5           &  1.7      &  25               & 4    & 2.8   &  0.22    &  2.87   \\
IG2                         & 3             &  1.5      &  9                  & 2    & --  &  0.14    &  1.62   \\
IG7                         & 3             &  1.5      &  11               & 2    & 0.9   &  0.23   &  2.13 \\
\hline
      \end{tabular}
\tablefoot{We adopt $M_{\rm BH}=3$~$M _{\odot}$, thus parameters are $R_{\rm S}=8.85\times10^5$~cm and $L_{\rm Edd} = 6\times10^{38}$~erg~s$^{-1}$ (Eddington limit for pure helium).
We assumed that $T_{\rm dbb}$ is equal to the temperature of the injected blackbody photons everywhere except for group IG2 where no seed disk photons were considered. $L_{\rm hot}$ is the luminosity of the hot flow;  it consists of synchrotron, bremsstrahlung, and Comptonized emission.
} 
\end{table*}

We also account for an additional source of seed photons in the calculations.
The  blackbody photons of temperature $T_{\rm dbb}$ and luminosity $L_{\rm d}$ (found from phenomenological fitting) are 
injected homogeneously into the medium.
After a steady-state solution is found, the remaining (not scattered) photons are then removed and an accretion disk spectrum with temperature \mbox{dependence} $T(R)\propto T_{\rm dbb} R^{-3/4}$ is added to the total spectrum instead of the blackbody spectrum.
We also compute the irradiated accretion disk contribution, which has temperature distribution 
$\displaystyle T_{\rm irr}(R) = T_{\rm irr, out}(R/R_{\rm out})^{-3/7}$ \citep{Cun76, FKR02}.
The outer disk temperature is calculated as
\begin{equation}
 T_{\rm irr, out} = \left(\frac{\alpha L_{\rm 1keV}}{4 \pi \sigma_{\rm SB} R_{\rm out}^2}\right)^{1/4},
\end{equation}
where $\sigma_{\rm SB}$ is the Stephan-Boltzmann constant;
$L_{\rm 1keV}$ is the ($\nu L_{\nu}$) luminosity at 1~keV (which is mostly absorbed and reprocessed in the accretion disk, 
see \citealt{PVR14}); and $\alpha$ is a factor accounting for the albedo, geometrical effects, and the bolometric correction (see, e.g., \citealt{SMM99}).

\subsection{Results}

The results of our simulations are shown in Fig.~\ref{fig:simulations}.
The observed fluxes are converted to luminosities assuming a distance of $D=3$~kpc.
The de-reddened {\it Swift}/UVOT fluxes are determined with the \citet{Fitzpatrick99} extinction curve (extinction parameter $R_V=3.1$) for two representative cases of reddening: we use $E(B-V)=0.45$ \citep{FMF14, RTC15} for our modeling, but also show the fluxes for $E(B-V)=0.34$ \citep{Cadolle07} to highlight how sensitive the overall spectral energy distribution is to this measurement.
Given the large systematic uncertainties in the reddening determination, we apply 10~percent flux errors to the UV data.
The spectra obtained from the simulations (dashed lines) complemented by contribution of the irradiated accretion disk (dot-dashed lines) are compared to the quasi-simultaneous UV data (yellow points) and the absorption corrected X-ray data obtained from the best fitting phenomenological X-ray model {\sc diskbb + compps}.
The irradiated disk flux drops towards the outburst tail.
We assume that these changes are entirely caused by the drop in the irradiated X-ray flux ($\nu L_{\nu}$ at 1~keV).
The other irradiated disk parameters are fixed to $\alpha=9\times10^{-2}$ and $R_{\rm out}=1.4\times10^{10}$~cm for all  four groups.
The index of the injected power-law electrons, $\Gamma_{\rm inj}=3.0$, was also assumed to be constant for every spectrum.
The other parameters are listed in Table~\ref{tab_simulat}.

The P03 observation displayed the softest spectrum of the ones we considered.
As can be seen from Fig.~\ref{fig:simulations}a, the spectrum can be  reproduced well in the disk Comptonization scenario.
The physical parameters that we obtained in the self-consistent calculations are, however, somewhat different from those found in the phenomenological data fitting.
The origin of these differences is likely related to the geometry and to the approximations made (such as escape probability formalism used in numerical modeling).
The electron temperature is the lowest of the cases considered, likely because the higher external soft photon luminosity from the disk cools the electrons more efficiently in the Compton up-scattering processes.
We find that the X-ray spectrum is concave, unlike in the phenomenological models.
There are two reasons: some curvature is induced by the bremsstrahlung radiation, but  a significant contribution also comes from the SSC~emission.
Compared to the disk Comptonization emission, the SSC~emission is much harder and its contribution can be estimated by extrapolating the low-energy part of spectrum to higher energies (dashed line at energies $\sim$0.01--0.1~keV).
The role of bremsstrahlung emission in total spectrum is less clear, however, as its strength depends on the absolute size of the medium.
It is therefore essential to know whether the emitting medium is one homogeneous blob (as assumed)
or whether it is composed of multiple blobs of smaller size, in which case the bremsstrahlung contribution would be smaller.

The disk spectrum is less prominent during the decline phase of the outburst (P18, Fig.~\ref{fig:simulations}b).
Simultaneously we see an \mbox{increased} spectral hardness and a slightly higher electron temperature.
These changes are consistent with the picture where a decreasing amount of soft disk photons results in a reduced electron cooling, and thus higher equilibrium electron temperature.
This observation likely marks the onset where the falling mass accretion rate causes the cold accretion disk to start receding from the BH and simultaneously reducing the electron number density in the hot flow.
The required parameter changes in our simulations with respect to the P03 observation is in agreement with this scenario: while the size $R$ must be increased, the optical depth and the disk seed photon luminosity drops.
Interestingly, the resulting bolometric luminosity of the hot medium is practically the same as in P03, only the disk luminosity changes.
Therefore, the drop in the accretion rate had to be compensated by increasing the hot flow size.

The spectrum during epoch IG02 (Fig.~\ref{fig:simulations}c) is the hardest out of the four cases considered. 
The accretion disk contribution is only marginally significant and, indeed, the disk seed photons are no longer required to reproduce the X-ray spectrum.
This may be interpreted as further truncation of cold accretion disk and development of a larger hot flow inside it.
The parameters in our simulations confirm this evolution: the hot flow size had to be increased even further and the optical depth had to be decreased along with the luminosity of the hot flow and the average magnetic field.
As noted by \citet{CDSG10}, the \swift/UVOT data lies on the continuation of the X-ray power law exactly as expected in the pure SSC scenario, but only if reddening is $E(B-V)=0.34$.
However, for the higher reddening $E(B-V)=0.45$ found by \citet{FMF14}, a much more substantial irradiated disk contribution is required in the optical/UV range.
 
In the spectrum of epoch IG07 (Fig.~\ref{fig:simulations}d), taken three years after IG02, we find that the disk had become hotter and the  temperature was close to that observed at the outburst peak.
At the same time, the spectral hardness and the reflection amplitude was similar to  P18, while X-ray luminosity was significantly lower and the electron temperature was significantly higher.
Our simulations show that the magnetic field must be smaller than  P18 and P03, as the irradiated disk contribution alone is no longer \mbox{sufficient} to reproduce a correct slope.
The magnetic field might indeed decrease in response to the decreased mass accretion rate \citep{NIA03, YN14}.
We also note that the hot flow size remains roughly constant while the disk becomes hotter.
One possibility is that the cold disk at this stage might penetrate deeper into the hot medium.

\section{Discussion}

\subsection{ Source of seed photons}

Our results in Figs.~\ref{fig:parameters} and~\ref{fig:parametersCOMPPS} show that the parameter evolution can be separated into two groups.
During the outburst peak, the spectrum is soft ($\Gamma \simeq 1.85$), the Compton reflection amplitude is substantial ($\Omega/2\pi \simeq 0.25$), 
the electron temperature is modest $T_{\rm e} \simeq 70$~keV, and the disk-to-Comptonization flux ratio is ${\simeq}0.2$.
All these spectral parameters change when the flux decays below a critical limit of $F\sim\!1.5\times 10^{-8}\,\ergcms$.
The spectra after that are much harder ($\Gamma \simeq 1.65$), the Compton reflection amplitude diminishes ($\Omega/2\pi \simeq 0.15$), 
the electron temperature doubles to $T_{\rm e} \simeq 140$~keV, and disk-to-Comptonization ratio drops  to ${\simeq}0.05$ or  even becomes undetectable.

We conclude that the observed parameter changes can be caused by an increase in the truncation radius (and a formation of a larger hot inner flow inside it), which in turn causes the main seed photon source for Comptonization to change from external disk emission to synchrotron emission from the hot flow.
The behavior of \swj\ is therefore similar to two other BH-LMXBs GX 339--4 and GRO J1655--40 \citep{SPDM11}, and to a sample of AGNs where this phenomenon seems to occur \citep{VVP11}.  

The decrease of the Compton reflection amplitude during the \swj\ outburst is similar to many other BH systems.
This is a well-known sign that the reflector, i.e., the thin disk, truncates during the outburst \citep{Gilfanov10}.
The increase electron temperature from $T_{\rm e}\sim 70$ to ${\sim}140$~keV can be seen as yet another consequence of a larger disk truncation.
Larger truncation results in a smaller number of seed photons entering the Comptonized region, which in turn leads to less efficient cooling of the electrons and thus higher equilibrium temperatures.
We speculate that a similar process causes the $T_{\rm e}$ variations seen during spectral state transitions of GX 339--4 \citep{MBH09}, the clear difference is that in \swj\ the transition happens during the hard state, and is not related to a state \mbox{transition}.

The scaling relations from \citet{B99ASP}, shown in Fig.~\ref{fig:parameters}, provide us with a further physical interpretation of the observed parameter dependencies within the hot flow paradigm.
The scaling in Eq.~(\ref{eq:ratio_theor}) is essentially an upper limit; i.e., for a given ratio of disk-to-Comptonization fluxes it gives the \mbox{softest} possible photon index $\Gamma$, or, equivalently, it gives the photon index under the assumption that the entire seed photon luminosity is intercepted by the hot medium.
As mentioned earlier, if we assume that even the hardest spectra are still produced by disk Comptonization, the inclination should be ${\sim}80\degr$, which we find implausible for two reasons:
(1) we do not see any dips or eclipses in the light curves; and
(2) such high inclination would require a NS accretor (see \citealt{NVP14}), which is incompatible with most of the spectral and timing characteristics of the source.
On the other hand, the hardest spectra are fully consistent $\delta=1/10$ tracks for the assumed inclination of $i=40\degr$.
If the seed photons are instead supplied by the synchrotron mechanism, the peak of the synchrotron spectrum should be located at the turn-over frequency $\nu_{\rm t}\sim5$~eV.
We therefore find it likely that from P20 -- below the critical flux limit of $F\sim\!1.5\times 10^{-8}\,\ergcms$ -- until IG07 the seed photons are produced through synchrotron-self-Comptonization in the hot flow.
Only during the outburst peak above the critical flux limit does the thin accretion disk likely extend so close to the black hole that it produces the majority of the seed photons for Comptonization.

\subsection{Soft excess}

The variability of the disk component normalization, which is proportional to the inner disk radius, is clearly inconsistent with the inner hot flow scenario.
The inner disk radius, when derived from the {{\sc diskbb}} model normalization, suggests that the truncation radius decreases when the rest of the parameters seem to suggest the opposite.
This issue was extensively discussed by \citet{CDSG10}, who proposed two alternatives.
Either the {{\sc diskbb}} component tracks a strongly irradiated inner disk edge that is truncated at a few tens of gravitational radii from the black hole; when \citet{CDSG10} modeled the spectrum with an irradiated disk model the inner radius increased towards the end of the outburst,
or the {{\sc diskbb}} component tracked a small residual inner disk (or a ring) located at ISCO with the hot flow located between this residual disk and the truncation radius much farther out.
With either of these alternatives, or even by assuming that the disk is always at the ISCO, we find it hard to explain the observations presented here. 
In particular, it is not clear what causes
(1) the inner disk temperature to remain constant at $T_{\rm dbb} \simeq 0.25$~keV while the disk flux changes by more than an order of magnitude;
(2) the {{\sc diskbb}} and {{\sc compps}} model normalizations to be nearly identical in the outburst peak; and
(3) the {{\sc diskbb}} normalizations in the 2012 soft state incursion to be very similar to the late outburst peak in 2005.
The first point in particular is challenging to tackle in any accretion flow geometry.
The second point, the one-to-one dependence between {{\sc diskbb}} and {{\sc compps}} model normalizations, suggests that about one-half of the disk flux is intercepted by the Comptonized region.
This may be a sign that the corona is patchy, i.e., the covering fraction of the disk by the corona is roughly 50~percent in the outburst peak. 
Alternatively, if the $T_{\rm dbb} \simeq 0.25$~keV disk is truncated hundreds of kilometers away from the vicinity of the black hole where most of the energy is released, 
this may also be a sign that the Comptonized region covers only a small fraction of the sky from the point of view of the disk (i.e., ``sombrero'' geometry; see \citealt{PKR97}).
On the other hand, in some spectra where no additional {{\sc diskbb}} component is detected (such as the IG07 in the outburst tail), the corona could cover the entire inner disk.
If this is the case, the corona or the hot flow would not receive the same disk flux that we measure at different outburst stages, and thus the relation between the spectral hardness and the seed photon flux in Eq.~(1) would be affected.
The third point, regarding a preferred size-scale, is puzzling given the evolution of other spectral parameters. 
If it is set by the disk edge being at the ISCO, then we cannot explain why the reflection amplitude, hardness, $F_{\rm dbb}/F_{\rm c}$ ratio, and timing properties (see below) evolve in the way observed.

\subsection{Simulations}

The parameter evolution in the self-consistent simulations is consistent with the scenario where the truncated disk recedes as the outburst declines (P03, P18, IG02).
The mass accretion rate drops and leads to an increase in the hot flow size (inside the truncation radius) and a decrease in the X-ray luminosity, 
electron number density, magnetic field, and the disk seed photon luminosity all support this view.
The increase in disk photon temperature with an apparent increase in the X-ray luminosity in IG07 might be connected to the increase in the 
mass accretion rate and penetration/covering of the disk into/by the hot flow. 

In particular, at the outburst peak, the extrapolation of the X-ray power law to lower energies considerably overpredicts the UV flux and at the same time the standard disk contribution appears to be insufficient.
By assuming that optical and UV wavelengths are dominated by the same component, both spectral \citep{CDSG10} and timing properties \citep{HBM09} can be explained by irradiation of the X-ray photons in outer parts of the accretion disk.
In the outburst tail, however, the picture is completely different.
By the time the X-ray flux dropped by more than an order of magnitude compared to the peak, the UV flux was reduced  by only a factor of two.
Furthermore, the UV spectrum in IG02 became significantly softer (Fig.~\ref{fig:simulations}, panel~c).
Here, however, the conclusions rely strongly on the uncertain reddening value for \swj.
As initially speculated by \citet{CDSG10}, the UV spectrum can be easily extrapolated from the X-ray power law for the reddening value obtained by \citet{Cadolle07}, suggesting that both of them  come from the same region. 
Our simulations show that the entire optical to hard X-ray spectrum  in this case can be produced by a single SSC emission region, meaning that the irradiated disk would not be required to describe data.
However, the higher reddening value $E(B-V)=0.45$ derived by \citet{FMF14} and \citet{RTC15} suggests a much steeper UV slope than  the X-rays.
For this higher reddening value, the spectrum can  still be adequately modeled, but the contribution from the irradiated disk flux in this case is instead comparable to the SSC-emission in the UV/optical regime during the IG02 spectrum.
The uncertainty in the reddening value thus limits the robustness of the simulations.
For the $E(B-V)=0.34$ case a smaller albedo factor and a higher outer disk radius would also provide a similar broad band spectral shape.
In order to be able to provide meaningful constraints for the parameters from our simulations, it would be necessary to know these two parameters using other means.
Broader energy coverage down to infrared frequencies would be helpful to constrain the spectral slopes, although  the emission from the jet and the companion star may start playing an increasingly important role there \citep{RTC15, TRK15}.

\subsection{Clues from X-ray timing analyses}

Some timing properties of \swj\ during the outburst support the picture where the disk geometry and the seed photon source change. 
On the one hand, the evolution of the low-frequency quasi-periodic oscillations (QPO) imply that the truncation radius increases during the outburst \citep{CDSG10}, but only if these QPOs track the truncation radius (see, e.g., \citealt{HWvdK01}).
Another piece of supporting evidence comes from the changes in the optical/\mbox{X-ray} cross-correlation functions (CCF) during the outburst.
During the outburst peak the CCF observed by \citet{HBM09} shows a simple positive lag, which is consistent with the reprocessing of the X-ray photons in the outer accretion disk.
However, in the tail, right after our spectral analysis suggests that the seed photon source changes, the CCFs show a completely different structure \citep{HBM09}.
The CCFs are now peculiar and feature an anti-correlation dip and two pronounced peaks, all of them first appearing at positive lags.
Similar CCF shapes are seen in summer 2007 and autumn 2008 (until IG07; \mbox{\citealt{DGS08,DSG11}}), but now the anti-correlation is seen at negative lags, while the correlation in still seen at positive lags (optical lags behind the X-rays).
These ``pre-cognition dips'' and the complex CCF shapes can be understood in the SSC-model as a combination of two effects (see \citealt{VPV11}, \citealt{PV14}, Fig.~13b and \citealt{VRD15}):
(1) the anti-correlation arises from an increased self-absorption in the optical synchrotron emission in response to a increase in mass accretion rate (i.e., X-ray flare) and 
(2) the same X-ray flare then irradiates the outer (truncated) disk, causing the second CCF component that is correlated at positive lags.
These CCF properties tie in nicely with our spectral simulations.
Initially, when almost all optical-UV emission seems to come from the irradiated outer disk, the CCFs are consistent with simple reprocessing, whereas in the outburst tail, when the SSC-emission starts to become the dominant source of optical emission, the complex CCF shape can also be  described with the same \mbox{SSC-model}.

\section{Summary and conclusions}

In this paper, we have reported on the evolution of spectral properties of the BH-LMXB \swj.
We find several new dependencies between various spectral components that help us to place constraints on the accretion disk geometry 
and the origin of seed photons for Comptonization throughout the ongoing outburst.   

We were able to determine that around a critical flux limit of $F\sim\!1.5\times 10^{-8}\,\ergcms$, \swj\ displays two distinct types of 
behavior.
Above the critical limit, the spectrum is softer ($\Gamma \simeq 1.85$), the Compton reflection amplitude is higher ($\Omega/2\pi \simeq 0.25$), 
the electron temperature is modest $T_{\rm e} \simeq 70$~keV, and the disk-to-Comptonization flux ratio is large.
Below this limit, the spectra are much harder ($\Gamma \simeq 1.65$), the \mbox{Compton} reflection amplitude diminishes ($\Omega/2\pi \simeq 0.15$), 
the electron temperature doubles to $T_{\rm e} \simeq 140$~keV, and the disk-to-Comptonization ratio drops or  even becomes undetectable.
Furthermore, our {{\sc compps}} analysis suggests that the hardening of the spectrum has two distinct origins: above the critical limit the spectrum hardens because the optical depth of the Comptonized region increases, whereas below the critical limit the optical depth drops substantially, but the spectrum nevertheless hardens further thanks to the significantly higher electron \mbox{temperature}.

We were also able to determine that
(1) the hard X-ray spectrum seen by \inte/ISGRI is strongly variable during the outburst tail, i.e., the cutoff energy varies from year to year with no clear dependency on other parameters such as the emitted flux; 
(2) there is a one-to-one relation between the {{\sc diskbb}} and {{\sc compps}} model normalizations during the outburst peak, implying that the Comptonized region intercepts roughly 50~percent of the disk emission;
(3) thermal Comptonization models fail to describe the spectral evolution during the short incursions to the soft state after 2010, suggesting that the Comptonization is more likely non-thermal in this state; and 
(4) there is a constant size scale of the soft X-ray excess during the entire outburst, which may be related to the innermost stable circular orbit around the black hole.

We find that the observed spectral changes can be understood in the truncated cold disk--inner hot flow scenario. 
In our view a substantial increase in the truncation radius is the main driver of the observed spectral variations.
Our phenomenological \mbox{X-ray} spectral analysis, and the optical-to-X-ray spectral simulations indicate that the geometrically thin, 
optically thick accretion disk is an adequate source of seed photons only at the peak of the outburst.
In the outburst tail the optical through \mbox{X-ray} spectral continuum can be better explained with a dominant contribution from SSC mechanism, which is favored by the harder \mbox{X-ray} spectral slope, the smaller amount of reflection, the QPO frequency variations, and  by the complex shape of the optical/X-ray cross-correlation function.

The main unresolved issue is related to the variation of the soft excess, which is likely produced by the accretion disk; the constant 
temperature of $T_{\rm dbb} \simeq 0.25$~keV over an order of magnitude changes in emitted flux, and the long-term variations are not 
understood. 
Further analysis and observations in soft X-rays are warranted to understand the observed trends.

\vspace*{2mm}
\begin{acknowledgement}
We would like to thank the referee for a very constructive report that helped to improve the manuscript, and we would also like to thank Sara Motta, Chris Done, and Juri Poutanen for helpful suggestions.
J.J.E.K. acknowledges support from the ESA research fellowship programme, the Emil Aaltonen Foundation, and the V\"{a}is\"{a}l\"{a} Foundation. 
AV acknowledges Academy of Finland grant 268740.
S.T. thanks the Russian Scientific Foundation for the support (grant 14-12-01287).
This research made use of the NASA Astrophysics Data System and of the data obtained from the High Energy Astrophysics Science Archive (HEASARC), 
which is a service of the Astrophysics Science Division at NASA/GSFC and the High Energy Astrophysics Division of the Smithsonian Astrophysical Observatory.
This work made use of data supplied by the UK Swift Science Data Centre at the University of Leicester.
This research has made use of MAXI data provided by RIKEN, JAXA and the MAXI team.
\end{acknowledgement}

\Online

\begin{appendix}

\onecolumn

\begin{table*}
\section{Best fitting model parameters}
\centering
\caption{\label{tab:nthComp}Best fitting parameters for fitting the X-ray data with a {{\sc constant} $\times$ {\sc wabs} $\times$ ({\sc diskbb} $+$ {\sc relf} $\ast$ {\sc nthcomp})} model.} 
\tiny
\renewcommand{\arraystretch}{1.16}
\begin{tabular}{@{}lcccccccr}
\hline\hline\noalign{\smallskip}
ID  & $kT_{\rm dbb}$                    & $N_{\rm dbb}$                 & $F_{\rm dbb}$                   & $\Gamma$                      & $\Omega/2\pi$                 & $N_{\rm nthC}$                        & $F_{\rm c}$                   & $\chi^{2}/\nu$ \\
\hline\noalign{\smallskip}
\multicolumn{9}{c}{Outburst peak: $kT_{\rm dbb} = kT_{\rm seed}$} \\
\hline\noalign{\smallskip}
P00 & $0.226_{-0.005}^{+0.005}$ & $44\,000_{-4000}^{+5000}$       & $0.246_{-0.007}^{+0.007}$         & $1.791_{-0.007}^{+0.007}$     & $0.22_{-0.03}^{+0.03}$        & $0.750_{-0.009}^{+0.009}$       & $1.50_{-0.02}^{+0.02}$        &  781/707 \\
P03 & $0.241_{-0.006}^{+0.007}$         & $64\,000_{-7000}^{+9000}$    & $0.47_{-0.02}^{+0.02}$         & $1.866_{-0.005}^{+0.005}$     & $0.30_{-0.02}^{+0.02}$        & $1.452_{-0.014}^{+0.014}$       & $2.28_{-0.02}^{+0.02}$        &  534/569 \\
P04 & $0.241_{-0.005}^{+0.005}$         & $56\,000_{-5000}^{+5000}$       & $0.411_{-0.010}^{+0.010}$       & $1.875_{-0.005}^{+0.005}$     & $0.25_{-0.02}^{+0.02}$         & $1.500_{-0.014}^{+0.014}$     & $2.29_{-0.02}^{+0.02}$        &  804/745 \\
P07 & $0.262_{-0.006}^{+0.006}$         & $41\,000_{-4000}^{+4000}$       & $0.416_{-0.010}^{+0.010}$       & $1.834_{-0.005}^{+0.005}$     & $0.25_{-0.02}^{+0.02}$         & $1.308_{-0.013}^{+0.013}$     & $2.31_{-0.02}^{+0.02}$        &  724/674 \\
P06 & $0.237_{-0.004}^{+0.004}$         & $61\,000_{-4000}^{+5000}$       & $0.418_{-0.008}^{+0.009}$       & $1.839_{-0.005}^{+0.005}$     & $0.26_{-0.02}^{+0.02}$         & $1.348_{-0.012}^{+0.012}$     & $2.31_{-0.02}^{+0.02}$        &  821/798 \\
P10 & $0.251_{-0.005}^{+0.005}$         & $48\,000_{-4000}^{+5000}$       & $0.409_{-0.010}^{+0.010}$       & $1.825_{-0.004}^{+0.004}$     & $0.246_{-0.014}^{+0.015}$         & $1.290_{-0.010}^{+0.010}$     & $2.33_{-0.02}^{+0.02}$        &  744/700 \\
P08 & $0.289_{-0.015}^{+0.014}$         & $27\,000_{-5000}^{+7000}$     & $0.40_{-0.03}^{+0.03}$                 & $1.818_{-0.004}^{+0.004}$     & $0.231_{-0.014}^{+0.014}$         & $1.23_{-0.02}^{+0.02}$        & $2.33_{-0.02}^{+0.02}$        &  286/312 \\
P09 & $0.239_{-0.004}^{+0.004}$         & $62\,000_{-4000}^{+5000}$       & $0.433_{-0.009}^{+0.009}$       & $1.827_{-0.004}^{+0.004}$     & $0.253_{-0.014}^{+0.014}$         & $1.309_{-0.009}^{+0.009}$     & $2.326_{-0.015}^{+0.015}$     &  848/776 \\
P11 & $0.24_{-0.02}^{+0.02}$            & $55\,000_{-14\,000}^{+20\,000}$    & $0.39_{-0.03}^{+0.03}$          & $1.818_{-0.005}^{+0.005}$     & $0.26_{-0.02}^{+0.02}$         & $1.27_{-0.02}^{+0.02}$        & $2.32_{-0.02}^{+0.02}$        &  360/335 \\
P12 & $0.222_{-0.004}^{+0.004}$         & $81\,000_{-7000}^{+7000}$    & $0.423_{-0.010}^{+0.010}$         & $1.825_{-0.004}^{+0.005}$     & $0.28_{-0.02}^{+0.02}$        & $1.297_{-0.011}^{+0.011}$       & $2.30_{-0.02}^{+0.02}$        &  805/771 \\
P13 & $0.235_{-0.005}^{+0.005}$         & $58\,000_{-6000}^{+7000}$     & $0.382_{-0.011}^{+0.011}$         & $1.822_{-0.005}^{+0.005}$     & $0.27_{-0.02}^{+0.02}$        & $1.278_{-0.011}^{+0.011}$       & $2.30_{-0.02}^{+0.02}$        &  675/685 \\
P15 & $0.227_{-0.004}^{+0.004}$         & $58\,000_{-5000}^{+6000}$       & $0.332_{-0.008}^{+0.008}$       & $1.805_{-0.005}^{+0.005}$     & $0.26_{-0.02}^{+0.02}$         & $1.192_{-0.010}^{+0.010}$     & $2.27_{-0.02}^{+0.02}$        &  842/791 \\
P16 & $0.250_{-0.006}^{+0.006}$         & $39\,000_{-4000}^{+5000}$       & $0.329_{-0.009}^{+0.010}$       & $1.800_{-0.005}^{+0.005}$     & $0.24_{-0.02}^{+0.02}$         & $1.162_{-0.011}^{+0.011}$     & $2.28_{-0.02}^{+0.02}$        &  702/664 \\
P18 & $0.230_{-0.011}^{+0.012}$         & $32\,000_{-7000}^{+9000}$       & $0.194_{-0.012}^{+0.012}$       & $1.740_{-0.005}^{+0.005}$     & $0.21_{-0.02}^{+0.02}$         & $0.758_{-0.009}^{+0.009}$     & $1.84_{-0.02}^{+0.02}$        &  411/448 \\
P19 & $0.226_{-0.007}^{+0.007}$         & $30\,000_{-4000}^{+5000}$       & $0.168_{-0.006}^{+0.007}$       & $1.726_{-0.005}^{+0.005}$     & $0.23_{-0.02}^{+0.02}$         & $0.675_{-0.007}^{+0.007}$     & $1.73_{-0.02}^{+0.02}$        &  653/662 \\
P20 & $0.231_{-0.014}^{+0.014}$         & $13\,000_{-3000}^{+4000}$       & $0.080_{-0.005}^{+0.006}$       & $1.675_{-0.006}^{+0.006}$     & $0.16_{-0.02}^{+0.02}$         & $0.425_{-0.006}^{+0.006}$     & $1.35_{-0.02}^{+0.02}$        &  454/476 \\
P21 & $0.22_{-0.02}^{+0.02}$            & $7000_{-2000}^{+3000}$        & $0.034_{-0.003}^{+0.003}$       & $1.655_{-0.012}^{+0.012}$     & $0.11_{-0.05}^{+0.05}$         & $0.248_{-0.006}^{+0.006}$     & $0.86_{-0.03}^{+0.03}$        &  518/538 \\
P23 & $0.15_{-0.02}^{+0.02}$            & $25\,000_{-11\,000}^{+25\,000}$     & $0.027_{-0.005}^{+0.008}$       & $1.658_{-0.009}^{+0.009}$     & $0.13_{-0.04}^{+0.04}$         & $0.210_{-0.003}^{+0.003}$     & $0.71_{-0.02}^{+0.02}$        &  610/658 \\
P24 & $0.14_{-0.02}^{+0.02}$            & $30\,000_{-20\,000}^{+40\,000}$     & $0.024_{-0.006}^{+0.011}$       & $1.673_{-0.008}^{+0.008}$     & $0.19_{-0.04}^{+0.04}$         & $0.194_{-0.003}^{+0.003}$     & $0.614_{-0.014}^{+0.014}$     &  622/647 \\
P26 & $[0.005]$                         & $...$                         & $...$                   & $1.655_{-0.010}^{+0.010}$     & $0.21_{-0.06}^{+0.06}$         & $0.132_{-0.002}^{+0.002}$     & $0.471_{-0.013}^{+0.014}$     &  608/583 \\
P30 & $[0.005]$                         & $...$                         & $...$                   & $1.663_{-0.009}^{+0.009}$     & $0.11_{-0.06}^{+0.06}$         & $0.124_{-0.002}^{+0.002}$     & $0.429_{-0.011}^{+0.012}$     &  562/609 \\
P31 & $[0.005]$                         & $...$                         & $...$                   & $1.641_{-0.010}^{+0.010}$     & $0.09_{-0.05}^{+0.05}$         & $0.115_{-0.002}^{+0.002}$     & $0.436_{-0.013}^{+0.013}$     &  619/523 \\
\noalign{\smallskip}\hline\noalign{\smallskip}
\multicolumn{9}{c}{INTEGRAL groups: $kT_{\rm dbb} = kT_{\rm seed}$} \\
\hline\noalign{\smallskip}
IG1 & $0.239_{-0.012}^{+0.013}$         & $9000_{-2000}^{+3000}$        & $0.067_{-0.004}^{+0.004}$    & $1.650_{-0.002}^{+0.002}$        & $0.114_{-0.009}^{+0.010}$     & $0.335_{-0.003}^{+0.003}$       & $1.200_{-0.008}^{+0.009}$     & 478/464 \\
IG2 & $[0.005]$                         & $...$                         & $...$                   & $1.639_{-0.004}^{+0.004}$     & $0.09_{-0.02}^{+0.02}$        & $0.1238_{-0.0007}^{+0.0008}$ & $0.475_{-0.005}^{+0.005}$        & 676/695         \\
IG4 & $0.15_{-0.02}^{+0.02}$            & $7000_{-3000}^{+10\,000}$       & $0.008_{-0.002}^{+0.003}$       & $1.652_{-0.004}^{+0.009}$     & $<0.10$                         & $0.103_{-0.002}^{+0.002}$     & $0.357_{-0.010}^{+0.010}$         & 641/638 \\
IG5 & $0.136_{-0.02}^{+0.015}$          & $40\,000_{-20\,000}^{+60\,000}$   & $0.028_{-0.007}^{+0.013}$         & $1.629_{-0.003}^{+0.003}$     & $0.077_{-0.010}^{+0.011}$     & $0.1226_{-0.0007}^{+0.0006}$    & $0.473_{-0.004}^{+0.004}$     & 525/535 \\
IG7 & $0.198_{-0.013}^{+0.013}$         & $9000_{-2000}^{+4000}$        & $0.031_{-0.002}^{+0.003}$       & $1.751_{-0.005}^{+0.005}$     & $0.19_{-0.02}^{+0.02}$         & $0.304_{-0.004}^{+0.004}$     & $0.699_{-0.008}^{+0.008}$     & 607/634 \\
IG13& $[0.005]$                         & $...$                         & $...$                   & $1.713_{-0.008}^{+0.008}$     & $0.11_{-0.03}^{+0.03}$        & $0.1229_{-0.0014}^{+0.0014}$    & $0.350_{-0.005}^{+0.006}$     & 411/414 \\
\noalign{\smallskip}\hline\noalign{\smallskip}
\multicolumn{9}{c}{ Failed transition of 2010} \\
\hline\noalign{\smallskip}
FT08 & $0.269_{-0.003}^{+0.003}$        & $...$                         & $...$                   & $3.61_{-0.02}^{+0.03}$        & $[0.11]$                      & $1.781_{-0.009}^{+0.009}$       & $0.810_{-0.006}^{+0.006}$     & 512/418 \\
FT09 & $0.390_{-0.006}^{+0.005}$        & $7900_{-200}^{+200}$          & $0.40_{-0.02}^{+0.02}$          & $2.75_{-0.11}^{+0.11}$                & $[0.11]$                        & $0.45_{-0.05}^{+0.06}$        & $0.23_{-0.02}^{+0.02}$         & 555/433 \\
FT10 & $0.411_{-0.006}^{+0.005}$        & $6300_{-200}^{+200}$          & $0.39_{-0.02}^{+0.02}$          & $2.75_{-0.11}^{+0.11}$                & $[0.11]$                        & $0.42_{-0.05}^{+0.06}$        & $0.21_{-0.02}^{+0.02}$         & 533/442 \\
FT11 & $0.292_{-0.003}^{+0.003}$        & $...$                         & $...$                   & $3.70_{-0.02}^{+0.03}$        & $[0.11]$                      & $1.681_{-0.008}^{+0.008}$       & $0.745_{-0.005}^{+0.005}$     & 477/446 \\
FT13 & $0.254_{-0.005}^{+0.005}$        & $...$                         & $...$                   & $3.53_{-0.03}^{+0.03}$        & $[0.11]$                      & $1.805_{-0.013}^{+0.013}$       & $0.839_{-0.011}^{+0.011}$     & 440/376 \\
FT14 & $0.298_{-0.006}^{+0.005}$        & $19\,000_{-900}^{+900}$         & $0.322_{-0.015}^{+0.014}$       & $2.67_{-0.06}^{+0.06}$        & $[0.11]$                         & $0.63_{-0.05}^{+0.05}$        & $0.31_{-0.02}^{+0.02}$         & 473/447 \\
FT17 & $0.247_{-0.002}^{+0.002}$        & $...$                         & $...$                   & $3.525_{-0.012}^{+0.013}$     & $[0.11]$                      & $1.639_{-0.006}^{+0.006}$       & $0.777_{-0.005}^{+0.005}$     & 776/590 \\
FT18 & $0.261_{-0.003}^{+0.003}$        & $...$                         & $...$                   & $3.59_{-0.02}^{+0.02}$        & $[0.11]$                      & $1.703_{-0.007}^{+0.007}$       & $0.783_{-0.006}^{+0.006}$     & 532/462 \\
\noalign{\smallskip}\hline\noalign{\smallskip}
\multicolumn{9}{c}{ Soft state of 2012} \\
\hline\noalign{\smallskip}
SS20 & $0.15_{-0.02}^{+0.02}$           & $28\,000_{-13\,000}^{+40\,000}$    & $0.029_{-0.005}^{+0.009}$       & $1.81_{-0.02}^{+0.02}$        & $[0.11]$                         & $0.179_{-0.005}^{+0.004}$     & $0.130_{-0.004}^{+0.004}$         & 467/493 \\
SS21 & $[0.005]$                        & $...$                         & $...$                   & $1.88_{-0.02}^{+0.02}$        & $[0.11]$                      & $0.206_{-0.003}^{+0.003}$       & $0.40_{-0.02}^{+0.03}$        & 247/266 \\
SS22 & $0.375_{-0.005}^{+0.005}$        & $10\,500_{-300}^{+300}$         & $0.45_{-0.02}^{+0.02}$          & $3.02_{-0.10}^{+0.10}$        & $[0.11]$                         & $0.59_{-0.06}^{+0.07}$        & $0.28_{-0.03}^{+0.03}$         & 324/400 \\
SS23 & $0.287_{-0.004}^{+0.004}$        & $...$                         & $...$                   & $3.89_{-0.03}^{+0.04}$        & $[0.11]$                      & $1.666_{-0.009}^{+0.009}$       & $0.739_{-0.006}^{+0.006}$     & 333/357 \\
SS24 & $0.303_{-0.004}^{+0.004}$        & $...$                         & $...$                   & $4.06_{-0.05}^{+0.05}$        & $[0.11]$                      & $1.912_{-0.012}^{+0.012}$       & $0.835_{-0.007}^{+0.007}$     & 278/325 \\
SS25 & $0.365_{-0.007}^{+0.006}$        & $11\,300_{-400}^{+400}$         & $0.43_{-0.03}^{+0.03}$          & $3.00_{-0.14}^{+0.14}$                & $[0.11]$                        & $0.59_{-0.09}^{+0.10}$                & $0.28_{-0.03}^{+0.04}$  & 368/341 \\
SS26 & $0.407_{-0.010}^{+0.009}$        & $7100_{-400}^{+400}$          & $0.42_{-0.05}^{+0.04}$          & $3.1_{-0.2}^{+0.2}$           & $[0.11]$                         & $0.54_{-0.12}^{+0.15}$                & $0.25_{-0.05}^{+0.06}$         & 237/265 \\
SS27 & $0.340_{-0.006}^{+0.005}$        & $19\,000_{-800}^{+800}$         & $0.54_{-0.03}^{+0.02}$          & $2.79_{-0.10}^{+0.11}$                & $[0.11]$                        & $0.68_{-0.12}^{+0.15}$        & $0.32_{-0.03}^{+0.04}$         & 307/323 \\
SS28 & $0.367_{-0.006}^{+0.005}$        & $7800_{-300}^{+300}$          & $0.305_{-0.012}^{+0.012}$       & $2.42_{-0.06}^{+0.06}$        & $[0.11]$                         & $0.47_{-0.04}^{+0.04}$        & $0.257_{-0.015}^{+0.02}$         & 396/429 \\
SS29 & $0.285_{-0.005}^{+0.005}$        & $16\,400_{-800}^{+900}$         & $0.232_{-0.009}^{+0.008}$       & $2.40_{-0.03}^{+0.03}$        & $[0.11]$                         & $0.60_{-0.03}^{+0.03}$        & $0.317_{-0.011}^{+0.012}$         & 468/486 \\
SS30 & $0.226_{-0.005}^{+0.005}$        & $38\,000_{-3000}^{+4000}$       & $0.214_{-0.007}^{+0.007}$       & $2.32_{-0.03}^{+0.03}$        & $[0.11]$                         & $0.55_{-0.02}^{+0.02}$        & $0.300_{-0.010}^{+0.011}$         & 448/455 \\
SS31 & $0.226_{-0.005}^{+0.005}$        & $32\,000_{-3000}^{+3000}$       & $0.184_{-0.005}^{+0.005}$       & $2.14_{-0.03}^{+0.03}$        & $[0.11]$                         & $0.419_{-0.015}^{+0.015}$     & $0.245_{-0.007}^{+0.007}$         & 446/467 \\
SS32 & $[0.005]$                        & $...$                         & $...$                   & $1.70_{-0.02}^{+0.02}$        & $[0.11]$                      & $0.100_{-0.002}^{+0.002}$       & $0.136_{-0.006}^{+0.007}$     & 187/205 \\
\hline 
\end{tabular}
\tablefoot{The absorption column is fixed to $0.2\times 10^{22}$~cm$^{-2}$, and the Compton reflection amplitude is from neutral, solar abundance material inclined at $40^\circ$ and it is given by the scaling factor $\Omega/2\pi$. 
The inner disk temperature is tied to the seed photon temperature for Comptonization in these fits.
The {{\sc diskbb}} and {\sc nthcomp} model component fluxes are unabsorbed and ``bolometric'' (computed with the {\sc cflux} convolution model) and they are given in units of $10^{-8}\,{\rm erg\,cm^{-2}\,s^{-1}}$.
The inner disk temperature $kT_{\rm dbb}$ is given in~keV.
We note that $F_{\rm c}$ does not contain the reflected flux.
}
\end{table*}

\begin{table*}
\centering
\caption{\label{tab:COMPPS}
Best fitting parameters for fitting the X-ray data with a {{\sc constant} $\times$ {\sc wabs} $\times$ ({\sc diskbb} $+$ {\sc relf} $\ast$ {\sc compps})} model.}
\tiny
\renewcommand{\arraystretch}{1.15}
\begin{tabular}{@{}lccccccccr}
\hline\hline\noalign{\smallskip}
ID              & $kT_{\rm seed}$       & $N_{\rm dbb}$                 & $F_{\rm dbb}$                   &$kT_{\rm e}$                   & $y$                           & $\Omega/2\pi$           & $N_{\rm cPS}$                         & $F_{\rm c}$                     & $\chi^{2}/\nu$ \\
\hline\noalign{\smallskip}
\multicolumn{10}{c}{Outburst peak: $kT_{\rm dbb} = kT_{\rm seed}$} \\
\hline\noalign{\smallskip}
P00 & $0.231_{-0.005}^{+0.005}$         & $34\,000_{-2000}^{+4000}$       & $0.208_{-0.008}^{+0.008}$       & $63_{-7}^{+6}$                & $1.079_{-0.012}^{+0.014}$         & $0.19_{-0.03}^{+0.03}$        & $30\,000_{-3000}^{+3000}$       & $1.43_{-0.03}^{+0.03}$  &  795/706 \\
P03 & $0.248_{-0.006}^{+0.007}$         & $46\,000_{-6000}^{+3000}$     & $0.38_{-0.02}^{+0.02}$         & $69_{-6}^{+5}$                & $0.953_{-0.010}^{+0.010}$     & $0.29_{-0.02}^{+0.02}$  & $49\,000_{-5000}^{+5000}$       & $2.27_{-0.02}^{+0.03}$         &  547/568 \\
P04 & $0.252_{-0.004}^{+0.005}$         & $34\,000_{-4000}^{+3000}$       & $0.301_{-0.015}^{+0.013}$       & $77_{-5}^{+6}$                & $0.929_{-0.012}^{+0.010}$         & $0.25_{-0.02}^{+0.02}$        & $49\,000_{-3000}^{+4000}$       & $2.33_{-0.03}^{+0.02}$  &  829/744 \\
P07 & $0.270_{-0.005}^{+0.006}$         & $29\,000_{-3000}^{+3000}$       & $0.335_{-0.014}^{+0.012}$       & $69_{-5}^{+7}$                & $1.003_{-0.012}^{+0.009}$         & $0.23_{-0.02}^{+0.02}$        & $32\,000_{-3000}^{+3000}$       & $2.27_{-0.03}^{+0.03}$  &  745/673 \\
P06 & $0.245_{-0.004}^{+0.004}$         & $43\,000_{-3000}^{+4000}$       & $0.340_{-0.012}^{+0.009}$       & $67_{-5}^{+5}$                & $0.997_{-0.009}^{+0.010}$         & $0.25_{-0.02}^{+0.02}$        & $46\,000_{-3000}^{+3000}$       & $2.26_{-0.03}^{+0.03}$  &  851/797 \\
P10 & $0.258_{-0.005}^{+0.005}$         & $34\,000_{-4000}^{+3000}$       & $0.331_{-0.010}^{+0.013}$       & $70_{-4}^{+6}$                & $1.018_{-0.010}^{+0.007}$         & $0.228_{-0.014}^{+0.02}$      & $37\,000_{-3000}^{+3000}$       & $2.29_{-0.02}^{+0.02}$  &  768/699 \\
P08 & $0.31_{-0.02}^{+0.02}$    & $17\,000_{-4000}^{+5000}$     & $0.32_{-0.03}^{+0.03}$         & $76_{-6}^{+7}$                & $1.018_{-0.013}^{+0.011}$     & $0.23_{-0.02}^{+0.02}$  & $19\,000_{-4000}^{+5000}$       & $2.30_{-0.03}^{+0.03}$         &  292/311 \\
P09 & $0.247_{-0.004}^{+0.004}$         & $44\,000_{-3000}^{+4000}$       & $0.354_{-0.012}^{+0.009}$       & $71_{-5}^{+4}$                & $1.013_{-0.008}^{+0.009}$         & $0.24_{-0.02}^{+0.02}$        & $44\,000_{-3000}^{+3000}$       & $2.29_{-0.02}^{+0.02}$  &  877/775 \\
P11 & $0.25_{-0.02}^{+0.02}$            & $37\,000_{-10\,000}^{+15\,000}$    & $0.30_{-0.03}^{+0.03}$          & $84_{-9}^{+9}$                & $1.011_{-0.015}^{+0.014}$         & $0.26_{-0.02}^{+0.02}$        & $44\,000_{-10\,000}^{+13\,000}$     & $2.32_{-0.04}^{+0.04}$  &  366/334 \\
P12 & $0.230_{-0.004}^{+0.004}$         & $56\,000_{-5000}^{+6000}$       & $0.340_{-0.02}^{+0.013}$        & $73_{-6}^{+5}$                & $1.013_{-0.005}^{+0.010}$         & $0.27_{-0.02}^{+0.02}$        & $57\,000_{-4000}^{+4000}$       & $2.28_{-0.03}^{+0.03}$  &  828/770 \\
P13 & $0.244_{-0.006}^{+0.006}$         & $39\,000_{-4000}^{+5000}$       & $0.301_{-0.015}^{+0.012}$       & $75_{-7}^{+6}$                & $1.016_{-0.011}^{+0.011}$         & $0.26_{-0.02}^{+0.02}$        & $45\,000_{-4000}^{+4000}$       & $2.28_{-0.03}^{+0.03}$  &  693/684 \\
P15 & $0.235_{-0.005}^{+0.005}$         & $41\,000_{-4000}^{+5000}$       & $0.266_{-0.009}^{+0.011}$       & $68_{-6}^{+5}$                & $1.051_{-0.009}^{+0.010}$         & $0.24_{-0.02}^{+0.02}$        & $47\,000_{-4000}^{+4000}$       & $2.20_{-0.03}^{+0.03}$  &  864/790 \\
P16 & $0.258_{-0.006}^{+0.006}$         & $27\,000_{-3000}^{+4000}$       & $0.260_{-0.006}^{+0.012}$       & $70_{-6}^{+6}$                & $1.058_{-0.010}^{+0.011}$         & $0.23_{-0.02}^{+0.02}$        & $33\,000_{-3000}^{+2000}$       & $2.21_{-0.04}^{+0.03}$  &  715/663 \\
P18 & $0.236_{-0.011}^{+0.012}$         & $22\,000_{-5000}^{+6000}$     & $0.150_{-0.012}^{+0.015}$         & $83_{-9}^{+9}$                & $1.163_{-0.012}^{+0.013}$     & $0.20_{-0.02}^{+0.02}$  & $30\,000_{-5000}^{+6000}$       & $1.79_{-0.04}^{+0.04}$         &  416/447 \\
P19 & $0.234_{-0.007}^{+0.008}$         & $21\,000_{-3000}^{+4000}$       & $0.133_{-0.008}^{+0.008}$       & $74_{-6}^{+9}$                & $1.199_{-0.013}^{+0.011}$         & $0.22_{-0.02}^{+0.03}$        & $26\,000_{-3000}^{+3000}$       & $1.64_{-0.35}^{+0.04}$  &  665/661 \\
P20 & $0.249_{-0.008}^{+0.02}$          & $6000_{-2000}^{+3000}$        & $0.053_{-0.009}^{+0.009}$       & $90_{-20}^{+20}$              & $1.30_{-0.03}^{+0.02}$         & $0.17_{-0.03}^{+0.04}$        & $14\,000_{-3000}^{+3000}$       & $1.30_{-0.06}^{+0.06}$  &  456/475 \\
P21 & $0.23_{-0.02}^{+0.02}$            & $4000_{-1500}^{+2000}$        & $0.024_{-0.005}^{+0.004}$       & $62_{-13}^{+20}$              & $1.36_{-0.03}^{+0.03}$         & $0.09_{-0.05}^{+0.05}$        & $9000_{-2000}^{+4000}$        & $0.76_{-0.04}^{+0.06}$  &  520/537 \\
P23 & $0.16_{-0.02}^{+0.02}$            & $12\,000_{-6000}^{+20\,000}$     & $0.017_{-0.003}^{+0.008}$       & $84_{-15}^{+20}$              & $1.35_{-0.02}^{+0.02}$         & $0.14_{-0.04}^{+0.05}$        & $31\,000_{-9000}^{+20\,000}$      & $0.66_{-0.03}^{+0.03}$  &  611/657 \\
P24 & $0.204_{-0.013}^{+0.013}$         & $...$                         & $...$                   & $122_{-14}^{+14}$             & $1.27_{-0.02}^{+0.02}$         & $0.26_{-0.04}^{+0.04}$        & $14\,000_{-3000}^{+4000}$       & $0.60_{-0.02}^{+0.02}$  &  633/647 \\
P26 & $[0.005]$                         & $...$                         & $...$                   & $170_{-90}^{+70}$             & $1.33_{-0.08}^{+0.06}$         & $0.23_{-0.07}^{+0.07}$        & $7.5_{-1.4}^{+1.1} \times 10^9$       & $0.51_{-0.08}^{+0.04}$  &  606/582 \\
P30 & $0.140_{-0.013}^{+0.013}$         & $...$                         & $...$                   & $120_{-30}^{+20}$             & $1.31_{-0.05}^{+0.04}$         & $0.16_{-0.07}^{+0.07}$        & $34\,000_{-11\,000}^{+15\,000}$     & $0.411_{-0.02}^{+0.015}$        &  555/607 \\
P31 & $0.188_{-0.015}^{+0.02}$          & $...$                         & $...$                   & $140_{-30}^{+20}$             & $1.40_{-0.02}^{+0.04}$         & $0.10_{-0.05}^{+0.06}$        & $10\,000_{-3000}^{+4000}$       & $0.44_{-0.02}^{+0.02}$  &  606/521 \\
\noalign{\smallskip}\hline\noalign{\smallskip}
\multicolumn{10}{c}{INTEGRAL groups: $kT_{\rm dbb} = kT_{\rm seed}$} \\
\hline\noalign{\smallskip}
IG1 & $0.255_{-0.015}^{+0.013}$         & $5400_{-700}^{+2000}$         & $0.050_{-0.005}^{+0.005}$       & $83_{-7}^{+7}$                & $1.369_{-0.006}^{+0.006}$         & $0.113_{-0.014}^{+0.012}$     & $9500_{-1500}^{+2000}$        & $1.11_{-0.02}^{+0.02}$  &  475/463 \\
IG2 & $0.104_{-0.010}^{+0.011}$         & $170\,000_{-11\,0000}^{+30\,0000}$ & $0.04_{-0.02}^{+0.04}$          & $165_{-14}^{+14}$             & $1.40_{-0.02}^{+0.02}$         & $0.10_{-0.02}^{+0.02}$        & $10\,0000_{-30\,000}^{+50\,000}$    & $0.508_{-0.013}^{+0.014}$       &  608/692 \\
IG4 & $0.188_{-0.015}^{+0.02}$          & $...$                         & $...$                   & $106_{-15}^{+13}$             & $1.31_{-0.02}^{+0.02}$         & $0.10_{-0.06}^{+0.06}$        & $9000_{-3000}^{+3000}$        & $0.348_{-0.014}^{+0.011}$       &  643/638 \\
IG5 & $0.224_{-0.010}^{+0.010}$         & $...$                         & $...$                   & $159_{-10}^{+12}$             & $1.384_{-0.015}^{+0.012}$         & $0.129_{-0.012}^{+0.014}$     & $6300_{-900}^{+1200}$ & $0.514_{-0.007}^{+0.010}$         &  524/535 \\
IG7 & $0.249_{-0.009}^{+0.009}$         & $...$                         & $...$                   & $124_{-6}^{+6}$               & $1.078_{-0.012}^{+0.012}$         & $0.25_{-0.02}^{+0.02}$        & $11\,000_{-1400}^{+2000}$       & $0.770_{-0.009}^{+0.008}$       &  611/634 \\
IG13& $0.34_{-0.03}^{+0.03}$            & $...$                         & $...$                   & $52_{-7}^{+8}$                & $1.21_{-0.02}^{+0.02}$         & $0.11_{-0.03}^{+0.04}$        & $1100_{-300}^{+500}$          & $0.281_{-0.011}^{+0.010}$       &  409/412 \\
\noalign{\smallskip}\hline\noalign{\smallskip}
\multicolumn{10}{c}{Failed transition of 2010: $kT_{\rm e} = 100$~keV, $kT_{\rm dbb} \neq kT_{\rm seed}$}              \\
ID   & $kT_{\rm dbb}$                   & $N_{\rm dbb}$                         & $F_{\rm dbb}$                   &$kT_{\rm seed}$                & $y$                           & $\Omega/2\pi$ & $N_{\rm cPS}$                   & $F_{\rm c}$                   & $\chi^{2}/\nu$ \\
\hline\noalign{\smallskip}
FT08 & $0.229_{-0.012}^{+0.012}$        & $49\,000_{-8000}^{+10\,000}$      & $0.29_{-0.02}^{+0.02}$          & $0.481_{-0.012}^{+0.014}$     & $0.143_{-0.005}^{+0.006}$         & $[0.11]$      & $3300_{-500}^{+500}$          & $0.44_{-0.02}^{+0.02}$         &  414/416 \\
FT09 & $0.20_{-0.02}^{+0.02}$           & $45\,000_{-13\,000}^{+20\,000}$             & $0.161_{-0.014}^{+0.014}$               & $0.452_{-0.007}^{+0.009}$     & $0.131_{-0.003}^{+0.004}$       & $[0.11]$      & $5000_{-500}^{+500}$          & $0.510_{-0.02}^{+0.015}$        &  467/432 \\
FT10 & $0.26_{-0.03}^{+0.04}$           & $14\,000_{-3000}^{+6000}$       & $0.13_{-0.03}^{+0.04}$                  & $0.472_{-0.012}^{+0.02}$      & $0.134_{-0.006}^{+0.011}$       & $[0.11]$      & $4000_{-900}^{+700}$          & $0.48_{-0.05}^{+0.03}$  &  497/441 \\
FT11 & $0.223_{-0.011}^{+0.011}$        & $50\,000_{-8000}^{+11\,000}$      & $0.268_{-0.012}^{+0.012}$       & $0.504_{-0.010}^{+0.011}$     & $0.122_{-0.004}^{+0.004}$         & $[0.11]$      & $2700_{-300}^{+300}$          & $0.42_{-0.02}^{+0.02}$         &  371/444 \\
FT13 & $0.237_{-0.014}^{+0.014}$        & $44\,000_{-6000}^{+9000}$       & $0.30_{-0.02}^{+0.02}$          & $0.475_{-0.015}^{+0.02}$      & $0.157_{-0.007}^{+0.009}$         & $[0.11]$      & $3400_{-700}^{+800}$  & $0.43_{-0.03}^{+0.03}$         &  404/374 \\
FT14 & $0.22_{-0.02}^{+0.02}$           & $52\,000_{-10\,000}^{+14\,000}$             & $0.28_{-0.03}^{+0.03}$                  & $0.43_{-0.02}^{+0.03}$        & $0.24_{-0.02}^{+0.02}$          & $[0.11]$      & $4000_{-1200}^{+1500}$        & $0.37_{-0.04}^{+0.04}$  &  405/446 \\
FT17 & $0.225_{-0.007}^{+0.007}$        & $58\,000_{-6000}^{+7000}$       & $0.324_{-0.009}^{+0.009}$       & $0.486_{-0.008}^{+0.008}$     & $0.151_{-0.003}^{+0.004}$         & $[0.11]$      & $2700_{-300}^{+300}$          & $0.377_{-0.012}^{+0.012}$         &  471/588 \\
FT18 & $0.237_{-0.009}^{+0.009} $       & $48\,000_{-6000}^{+7000}$       & $0.331_{-0.013}^{+0.013}$       & $0.510_{-0.012}^{+0.013}$     & $0.144_{-0.005}^{+0.005}$         & $[0.11]$      & $2300_{-300}^{+300}$          & $0.37_{-0.02}^{+0.02}$         &  425/460 \\
\noalign{\smallskip}\hline\noalign{\smallskip}
\multicolumn{10}{c}{Soft state of 2012: $kT_{\rm e} = 100$~keV, $kT_{\rm dbb} \neq kT_{\rm seed}  $}             \\
ID      & $kT_{\rm dbb}$                & $N_{\rm dbb}$                 & $F_{\rm dbb}$                   &$kT_{\rm seed}$                & $y$                           & $\Omega/2\pi$ & $N_{\rm cPS}$                           & $F_{\rm c}$                   & $\chi^{2}/\nu$ \\
\hline\noalign{\smallskip}
SS20 & $...$                            & $...$                         & $...$                   & $0.194_{-0.013}^{+0.013}$     & $0.92_{-0.02}^{+0.02}$         & $[0.11]$      & $18\,000_{-4000}^{+6000}$               & $0.139_{-0.002}^{+0.002}$         &  480/494 \\
SS21 & $...$                            & $...$                         & $...$                   & $0.21_{-0.03}^{+0.03}$        & $0.94_{-0.04}^{+0.04}$         & $[0.11]$      & $12\,000_{-5000}^{+9000}$               & $0.143_{-0.004}^{+0.005}$         &  236/265 \\
SS22 & $0.360_{-0.02}^{+0.012}$         & $14\,000_{-800}^{+1000}$        & $0.50_{-0.08}^{+0.04}$          & $0.65_{-0.10}^{+0.10}$        & $0.25_{-0.06}^{+0.05}$         & $[0.11]$      & $400_{-300}^{+600}$                   & $0.21_{-0.05}^{+0.09}$         &  320/399 \\
SS23 & $0.314_{-0.010}^{+0.010}$        & $19\,000_{-2000}^{+2000}$       & $0.40_{-0.02}^{+0.02}$          & $0.64_{-0.03}^{+0.04}$        & $0.138_{-0.011}^{+0.012}$         & $[0.11]$      & $500_{-200}^{+200}$                   & $0.22_{-0.02}^{+0.03}$         &  317/355 \\
SS24 & $0.31_{-0.02}^{+0.02}$           & $21\,000_{-2000}^{+3000}$       & $0.40_{-0.05}^{+0.04}$          & $0.57_{-0.04}^{+0.04}$        & $0.120_{-0.02}^{+0.012}$         & $[0.11]$      & $1200_{-500}^{+700}$                  & $0.30_{-0.05}^{+0.05}$         &  261/323 \\
SS25 & $0.33_{-0.02}^{+0.02}$           & $16\,700_{-1300}^{+2000}$       & $0.40_{-0.08}^{+0.06}$          & $0.54_{-0.05}^{+0.05}$        & $0.19_{-0.04}^{+0.04}$         & $[0.11]$      & $1400_{-700}^{+1300}$                 & $0.31_{-0.07}^{+0.08}$         &  349/340 \\
SS26 & $0.408_{-0.02}^{+0.014}$         & $8800_{-700}^{+900}$          & $0.52_{-0.09}^{+0.03}$          & $0.86_{-0.2}^{+0.3}$          & $0.28_{-0.08}^{+0.08}$                         & $[0.11]$      & $90_{-70}^{+500}$                     & $0.13_{-0.03}^{+0.10}$  &  236/264 \\
SS27 & $0.331_{-0.02}^{+0.013}$         & $24\,000_{-2000}^{+3000}$       & $0.62_{-0.13}^{+0.04}$          & $0.65_{-0.2}^{+0.15}$                 & $0.34_{-0.10}^{+0.06}$          & $[0.11]$      & $400_{-300}^{+2000}$                  & $0.22_{-0.05}^{+0.13}$  &  306/322 \\
SS28 & $0.380_{-0.008}^{+0.007}$        & $8800_{-500}^{+500}$          & $0.395_{-0.011}^{+0.008}$       & $1.1_{-0.2}^{+0.2}$           & $0.52_{-0.10}^{+0.07}$         & $[0.11]$      & $40_{-20}^{+30}$                      & $0.144_{-0.010}^{+0.013}$         &  396/428 \\
SS29 & $0.304_{-0.006}^{+0.006}$        & $18\,500_{-1200}^{+1300}$       & $0.341_{-0.006}^{+0.005}$       & $0.91_{-0.07}^{+0.07}$        & $0.53_{-0.04}^{+0.03}$         & $[0.11]$      & $90_{-20}^{+40}$                      & $0.187_{-0.008}^{+0.010}$         &  453/485 \\
SS30 & $0.241_{-0.007}^{+0.007}$        & $41\,000_{-4000}^{+4000}$       & $0.302_{-0.006}^{+0.006}$       & $0.70_{-0.06}^{+0.06}$        & $0.60_{-0.02}^{+0.02}$         & $[0.11]$      & $240_{-70}^{+110}$                    & $0.187_{-0.012}^{+0.02}$         &  499/454 \\
SS31 & $...$                            & $...$                         & $...$                   & $0.226_{-0.004}^{+0.004}$     & $0.361_{-0.005}^{+0.005}$         & $[0.11]$      & $49\,000_{-4000}^{+4000}$               & $0.410_{-0.006}^{+0.006}$         &  471/468 \\
SS32 & $...$                            & $...$                         & $...$                   & $[0.005]$                     & $1.25_{-0.09}^{+0.10}$         & $[0.11]$      & $6.1_{-0.7}^{+0.8} \times 10^9$       & $0.098_{-0.002}^{+0.002}$         &  187/205 \\
\noalign{\smallskip}\hline 
\end{tabular}
\tablefoot{The absorption column is fixed to $0.2\times 10^{22}$~cm$^{-2}$, and the Compton reflection amplitude is from neutral, solar abundance material inclined at $40^\circ$ and it is given by the scaling factor $\Omega/2\pi$. 
During the failed transition of 2010 and the 2012 soft state the inner disk temperature is not tied to the seed photon temperature for Comptonization, and the electron temperature is fixed to 100~keV.
The {{\sc diskbb}} and {\sc nthcomp} model component fluxes are unabsorbed and ``bolometric'' (computed with the {\sc cflux} convolution model) and they are given in units of $10^{-8}\,{\rm erg\,cm^{-2}\,s^{-1}}$.
The inner disk temperature $kT_{\rm dbb}$ and the seed photon temperature $kT_{\rm seed}$ are given in~keV.
We note that $F_{\rm c}$ does not contain the reflected flux.
}
\end{table*}

\end{appendix}

\label{lastpage}

\end{document}